\documentclass[12pt]{article}
\usepackage{amsmath}
\usepackage{amsfonts}
\usepackage{graphicx,psfrag,epsf}
\usepackage{enumerate}
\usepackage{algorithm}
\usepackage{algpseudocode}
\usepackage[round]{natbib}  
\usepackage{xcolor}
\usepackage{url} 
\usepackage{setspace}  
\usepackage[normalem]{ulem}
\usepackage{algorithm}
\usepackage{tikz}
\usepackage{hyperref}
\usetikzlibrary{positioning}
\usepackage{setspace}
\usepackage{booktabs}
\usepackage{multirow}
\usepackage{soul}
\usepackage{adjustbox}

\newcommand{\blind}{1}
\newcommand{\indep}{\perp \!\!\! \perp}

\newcommand{\citeay}[1]{\citeauthor{#1} \citeyear{#1}}

\newcommand{\directcite}[1]{\citeauthor{#1} (\citeyear{#1})}

\begin{document}

\def\spacingset#1{\renewcommand{\baselinestretch}%
{#1}\small\normalsize} \spacingset{1}

\spacingset{1.45}

\if1\blind
{
  \title{\bf 
   Evaluation of the HeartSteps Online Sampling Algorithm}
  \author{Xiang Meng, Walter Dempsey, Peng Liao, Nick Reid, Pedja Klasnja, Susan Murphy}
  \maketitle
} \fi

\if0\blind
{
  \bigskip
  \bigskip
  \bigskip
  \begin{center}
    {\LARGE\bf HeartSteps V2 and V3 Study: Design of Interventions and Analysis in Time-Varying Causal Excursion Effect}
\end{center}
  \medskip
} \fi

\begin{abstract}
Micro-randomized trials (MRTs), which sequentially randomize participants at multiple decision times, have gained prominence in digital intervention development. These sequential randomizations are often subject to certain constraints. In an MRT called HeartSteps V2V3, where an intervention is designed to interrupt sedentary behavior, two core design constraints need to be managed: an average of 1.5 interventions across days and the uniform delivery of interventions across decision times. Meeting both constraints, especially when the times allowed for randomization are not determined beforehand, is challenging. An online algorithm was implemented to meet these constraints in the HeartSteps V2V3 MRT. We present a case study using data from the HeartSteps V2V3 MRT, where we select appropriate metrics, discuss issues in making an accurate evaluation, and assess the algorithm’s performance. Our evaluation shows that the algorithm performed well in meeting the two constraints. Furthermore, we identify areas for improvement and provide recommendations for designers of MRTs that need to satisfy these core design constraints.
\end{abstract}

\bigskip

\section{Introduction}

Micro-randomized trials \citep[MRTs; ][]{dempsey2015randomised, klasnja2015microrandomized, liao2016sample} are experimental studies that sequentially randomize each participant at thousands of decision times over the course of a given trial. MRTs are frequently used in digital intervention development \citep{rabbi2018toward, bidargaddi2018prompt, kramer2019investigating, bell2020notifications, kroska2020optimizing, necamp2020assessing, liao2020personalized, nahum2021mobile, golbus2021microrandomized, dowling2022gambling, qian2022microrandomized, spruijt2022advancing, trella2022designing, govuk2023mrt, liu2023microrandomized, wang2023effectiveness}. Digital interventions involve the use of smart devices such as a smartphone, smartwatch, or other wearable device to deliver sequences of treatments  to individuals as they go about their daily lives. These treatments comprise messages with varying content, such as motivational messages, nudges of different types, reminders, and so on.

Multiple MRTs have been conducted in the process of developing the HeartSteps intervention package designed to encourage physical activity (\citeay{klasnja2019efficacy}, \citeay{liao2020personalized}, \citeay{spruijt2022advancing}). One intervention component in the package, called the anti-sedentary intervention, is the delivery of messages that are designed to interrupt sedentary behavior (to stand up, stretch, and/or move around).  
In the HeartSteps V2V3 MRTs \citep[HSV2V3; ][]{liao2020personalized, spruijt2022advancing}, decision times occur at five-minute intervals over the course of each study day. An individual who has taken fewer than 150 steps in the 40 minutes prior to a decision time is considered sedentary. Anti-sedentary messages were randomized to be sent or not at sedentary decision times throughout the day. For a decision time \( t \), we use \( B_t \) to denote sedentariness: \( B_t = 1 \) indicates that time \( t \) is a sedentary time, and \( B_t = 0 \) indicates otherwise.

The anti-sedentary intervention in the HSV2V3 MRTs needs to meet two common MRT design constraints: 
(1) the \emph{average treatment constraint}, which restricts the average number of treatments delivered to participants in a pre-specified time range, e.g. a four-hour time block in a study day. Such constraints can help reduce the burden to participants of digital interventions (\citeay{klasnja2008using}, \citeay{figueroa2021adaptive}, \citeay{liao2018just}); and (2) the \emph{uniformity constraint} aims to ensure that treatments are distributed uniformly across times allowed for randomization.
This is to ensure context sampling rates align with the frequency at which the contexts occur during sedentary times.
A context is information concerning an individual in which the treatment might have an effect (e.g., whether an individual is stressed or not) \citep{nahum2018just}. Since the distribution of the context is not known beforehand, treatments must be uniformly distributed across times allowed for randomization to ensure no context is favored. Otherwise, it is possible that the number of treatments delivered in certain context (e.g., when an individual is stressed) is low (when stressed sedentary times occur fairly frequently).  As a result, the ability to learn the causal effect in this context is harmed since we have little post-treatment data for that block \citep{liao2018just, dempsey2020stratified}.

Both design constraints can be satisfied by a straightforward MRT design (i.e., randomization probabilities) when the total number of sedentary times is known in a pre-specified time range.
HSV2V3 consists of 12-hour days segmented into three equal four-hour blocks, each comprising 48 decision times. The scientific team aims to ensure that on average for each participant, 0.5 interventions are delivered uniformly per four-hour block during sedentary times (\( B_t = 1 \)) with none happening during non-sedentary times (\( B_t = 0 \)). Suppose beforehand that 20 of 48 decision times are sedentary times. In this case, randomizing the ``send a message" option at each sedentary time with a probability of \(\frac{0.5}{20} = 0.025\) will ensure both constraints are met: The average treatment constraint is met because the treatment probabilities sum up to $0.5$, the desired average. The uniformity constraint is met because the treatment probabilities are equal to each other. However, the number of sedentary times in each block for each participant is completely unknown at the beginning of the block, so this strategy cannot be applied.

To address this challenge, an online algorithm called Sequential Risk Time Sampling (SeqRTS) was developed \citep{liao2018just, dempsey2020stratified}. In simulated data, SeqRTS was shown to meet the two constraints well on average across participants. However, the question of whether the algorithm can successfully meet the two constraints when implemented in the real world is still uncertain, as success on a simulated dataset does not guarantee success in the real world due to the existence of uncontrolled factors \citep{liu2022real, thompson2021replication}. A randomized trial is the gold standard for determining an algorithm's efficacy in mobile health and, in general, the entire healthcare sector when making real-time interventions \citep{lee2018effective}. SeqRTS was implemented in the HSV2V3 study to manage the randomization of anti-sedentary messages with the aim of meeting the two constraints. We conducted an evaluation using data collected from HSV2V3 to assess the algorithm's performance in meeting these constraints.

Our analysis shows that the algorithm successfully met both constraints on average across all participants in HSV2V3. We also show that it did not meet the ideal goal of achieving both constraints for each individual participant. This is unsurprising to the algorithm's designers because this is one of the first few studies in which this algorithm has been implemented in a real study. We identified participants and blocks where the algorithm underperformed and found baseline and time-varying variables that characterize them. After identifying these participants and blocks, we provide suggestions to move the algorithm's performance closer to the ideal goal of meeting both constraints for each individual participant.

SeqRTS is designed to perform well even with moderate effectiveness of the anti-sedentary messages. Therefore, if the messages result in mildly reduced sedentary behavior, the algorithm should still function effectively, and there is no need to adjust the evaluation to account for days with low sedentary behavior.

Our contributions are summarized as follows:
\begin{enumerate}
    \item \textbf{Evaluation of a Real-World Implementation of an Online Algorithm} We present a comprehensive study of 82 participants over 9,600 days, demonstrating that the SeqRTS algorithm performed well in the real-world setting of an MRT for meeting the design constraints. We select appropriate metrics and discuss conditions under which the evaluation is fair. This is the first extensive evaluation of such an algorithm in this context.  

    \item \textbf{Methodological Contributions:} While the number of interventions delivered is an obvious metric to assess SeqRTS's performance on achieving the average treatment constraint, the appropriate measure for evaluating the algorithm's performance on the uniformity constraint is less obvious. We discuss possible metrics for uniformity and propose using the Mean Absolute Deviation (MAD) as a suitable measure. The MAD has advantage over KL divergence, the uniformity metric suggested in previous work by \cite{liao2018just}.

    \item \textbf{Suggestions for Future Improvement of the Algorithm:} 
    We identify key variables that can improve the algorithm's performance closer to the ideal goal of meeting both constraints for each individual participant. Specifically, we identify variables correlated with SeqRTS's under-performance  and quantify their relationship to the algorithm's under-performance. We then discuss improvement strategies utilizing these findings, thereby offering actionable insights for further refinement of the algorithm.
\end{enumerate}

This study demonstrates the effectiveness of SeqRTS in meeting the two design constraints in a real-world setting and provides a methodological blueprint for future algorithmic advancements in similar domains.

The organization of the paper is as follows. In Section \ref{sec:HSV2V3-trial}, we introduce the HeartSteps V2V3 trial. In Section \ref{sec:SeqRTS}, we describe the SeqRTS algorithm used in this trial. In Section \ref{sec:evaluation}, we evaluate the algorithm's performance on the average treatment constraint and the uniformity constraint. In Section \ref{sec:covars}, we investigate baseline variables and time-varying covariates that could be used to improve the algorithm's performance, and propose improvement strategies for the algorithm. We conclude the paper by discussing limitations and future work in Section \ref{sec:conclusion}.

\subsection{Related Work}
The first category of related work pertains to the literature where sedentary times are of interest. Sedentary times are instances of broadly defined risk times, which are times of elevated stress or heightened risk for participants. Assessing treatment effects during risk times is of interest to many investigators. An example besides HSV2V3 is Sense2Stop \citep{battalio2021sense2stop}, a smoking cessation trial. In Sense2Stop, risk times are times when a participant (a smoker) is stressed. An investigator is interested in assessing the effect of a message to reduce the probability of smoking at risk times. Risk times are infrequent and occur in a stochastic manner. To handle the inherently random temporal distribution of risk times, \cite{dempsey2020stratified} defined the stratified micro-randomized trial (sMRT), where strata are times when a participant is at different levels of risk. In the anti-sedentary intervention of HSV2V3, risk times are sedentary times, and the number of strata is two (sedentary times and non-sedentary times). In general, sMRTs may encompass more than two strata.

The second category of related work focuses on existing evaluations of online algorithms. Online algorithms are designed for scenarios in which data or inputs are received incrementally over time \citep{zinkevich2003online}. In these circumstances, decisions are made in real-time as data point arrives, usually with incomplete knowledge or a limited understanding of potential future events. The decision problem that we have introduced here is an online problem because the arrival of future sedentary times is unknown.

Many works on evaluations of online algorithms exist. There are theoretical analyses that focus on comparing the performance of an online algorithm to that of an ideal offline counterpart with complete foresight \citep[competitive analysis; ][]{lin2012online, awerbuch1997buy}, or aim to establish theoretical bounds on the regret, the difference between the algorithm's performance and that of an optimal decision-making policy \citep{auer2008near, azar2017minimax, zhang2022online, arora2012online, lattimore2020bandit}. Our work differs from the above because it is an empirical evaluation.

Our work also differs from previous empirical evaluations. Empirical studies of online algorithms often rely on simulations to generate outcomes, even when using real-world datasets \citep{vermorel2005multi, agarwal2023empirical}.  Our evaluation differs because we evaluate the algorithm on the trial where it was implemented, 
Evaluating algorithms on real data is crucial for understanding their practical effectiveness and robustness in real-world scenarios \citep{burns2022real}. In real-world applications, online algorithms have been employed in studies like those by \cite{mate2022field} and \cite{verma2023restless}, which utilized online multi-armed bandit algorithms to make optimal allocation of service calls from health workers to prevent drops in partcipant's engagement in a healthcare study. They aimed to ensure that the best possible decisions were made with regard to an objective function. Our work has a different focus; we concentrate on ensuring that constraints are met.

The third category of related work focuses on the notion of uniformity in experimental design. In terms of managing the uniformity, the closest comparison in the literature is the biased coin design (BCD), first proposed by \directcite{efron1971forcing} and later extended by \directcite{wei1978adaptive}. The BCD appears in a two-arm (one treated, one control) randomized control trial (RCT) setting. In BCD, the notion of uniformity is the balance between the number of treated subjects and the number of controls. The biased coin design (BCD) aims to maintain a good balance by assigning a lower probability to the treated arm when the number of treatments exceeds the number of controls at the treatment assignment time, and vice versa.

The similarity between the balance in BCD and uniformity in SeqRTS is that they both tackle a decision problem where future event knowledge is limited (i.e., an online decision problem): in BCD, the future number of participants is unknown, and in SeqRTS, the future number of sedentary times is unknown. The key difference lies in the quantity that they aim to maintain well. The quantity that the BCD aims to maintain well is the ratio of treated participants to control participants. Keeping a good balance requires that this ratio is as close to one as possible. For SeqRTS, the quantity to maintain well is the variance of the treatment probabilities across the sedentary times (see Section \ref{sec:measures} for a detailed discussion). The uniformity constraint requires the variance to be as low as possible, ideally zero. Unlike BCD, it is not a concern if the number of treated times does not equal the number of control times.

\section{The HeartSteps V2V3 Trial}
\label{sec:HSV2V3-trial}

HeartSteps is a mobile intervention developed to encourage physical activity. HeartSteps has multiple intervention components (referred to as the ``package" in the Introduction). In this paper we focus on the  anti-sedentary intervention component that aims to disrupt the sedentary behavior by providing suggestions to stand up, stretch, and/or move around via the participant's smartphone. We use  a ``participant" or a ``user" interchangably to refer a person enrolled in the study. HeartSteps V2 and V3 (HSV2V3) involved participants who were newly diagnosed with  stage 1 hypertension (a diastolic pressure of 80 to 89 mm Hg or a systolic pressure of 130 to 139 mm Hg, \citeay{mayoclinic2021}).  The primary difference between HeartSteps V2 and HeartSteps V3 was the study duration. In V2, the duration of each participant's participation was 90 days, running from May 2019 to January 2020, whereas in V3, it extended to 270 days, beginning in October 2019 and ending in February 2021. Note the participants entered the study at various dates so the whole span of the study is longer than 90 days for V2 and 270 days for V3. 

In HeartSteps V2V3, all participants were provided a Fitbit tracker and the HeartSteps mobile phone application. Each intervention is delivered as a notification on the participant's smartphone via the HeartSteps phone app. We refer to these notifications as the treatment, and in the case of the anti-sedentary intervention, the treatment was an anti-sedentary message. Each participant was asked to wear the Fitbit tracker and to install HeartSteps software on their tracker, which we call the HeartSteps watch face.  The watch face was used to transmit step count data at short intervals in time to the HeartSteps cloud server. 

A baseline period is a period during which data is collected on the participant's activity without any intervention. The HeartSteps interventions start the day after the baseline period ends. The baseline period ends after 7 days of baseline information is collected. Sometimes it takes more than 7 days for some participants to finish the baseline period because participants may not wear the watch on some days. The median length of the baseline period is 11 days.

On each study day, the eligibility for anti-sedentary messages is calculated for a twelve-hour period whose end time is one hour after the participant's usual post-dinner time (collected before the beginning of the day). This eligibility is determined every five minutes, referred to as decision times. At each decision time, a participant was deemed either \lq\lq unavailable for treatment\rq\rq\ (see below for a definition) or ``available for treatment". If available for treatment, the participant was classified in one of two strata, currently sedentary or not currently sedentary, on the basis of the step count data over the last 40 minutes from the HeartSteps cloud server. Participants were considered sedentary if they had taken fewer than 150 steps in the prior 40 minutes. HeartSteps only randomized the delivery of the anti-sedentary messages at times when participants were both available and sedentary. We call these decision times \lq\lq risk times.\rq\rq\ Both availability and sedentariness at decision time \( t \) are determined right before time \( t \).
 
A participant can be unavailable for various reasons. First, participants are considered unavailable if they have  been active recently, that is, if they have taken more than 2,000 steps in the previous 120 minutes. Second, participants are considered unavailable if the watch face is not connected to the cloud server. This includes both decision times at which the  participant  is not wearing the Fitbit wrist tracker and decision times  when the participant is  wearing  the Fitbit but the watch face is unable to communicate with  the cloud based server. If the cloud based server does not have any records from the watch face for the 40 minutes prior to a decision time, the decision time is  defined as unavailable. Third, decision times are considered unavailable if the participant has been sent any type of notification (including an anti-sedentary message) in the prior hour.  Finally, participants were able to adjust settings on the HeartStep app to indicate that they do not want to be disturbed. All decision times occurring during these do not disturb times are considered unavailable. 

Unfortunately, when storing the data, only one binary availability indicator was kept for the aforementioned four criteria, instead of four separate binary variables indicating availability for each criterion. This data storage issue creates challenges when analyzing the algorithm's performance, primarily because the third criterion causes availability to depend on the algorithm implemented in the trial. We discuss the implications of this issue in Section 6.

The two HeartStep studies recruited 113 adults;  19 registered but did not participate with the result that 94 participants participated in the combined HSV2V3 study. Data from a further 12  participants were excluded due to technological issues.  In particular, 9 of 94 participants (9.6\%) are excluded due to the watch face not getting set up correctly, as it was not possible to determine whether a decision time was a risk time (with participant both available and sedentary). 3 of 94 participants (3.2\%) are excluded due to no Fitbit data. In summary, we use the data from  38 participants in V2 and 44 participants in  V3.

\begin{figure}
    \centering
    \includegraphics[width=\linewidth]{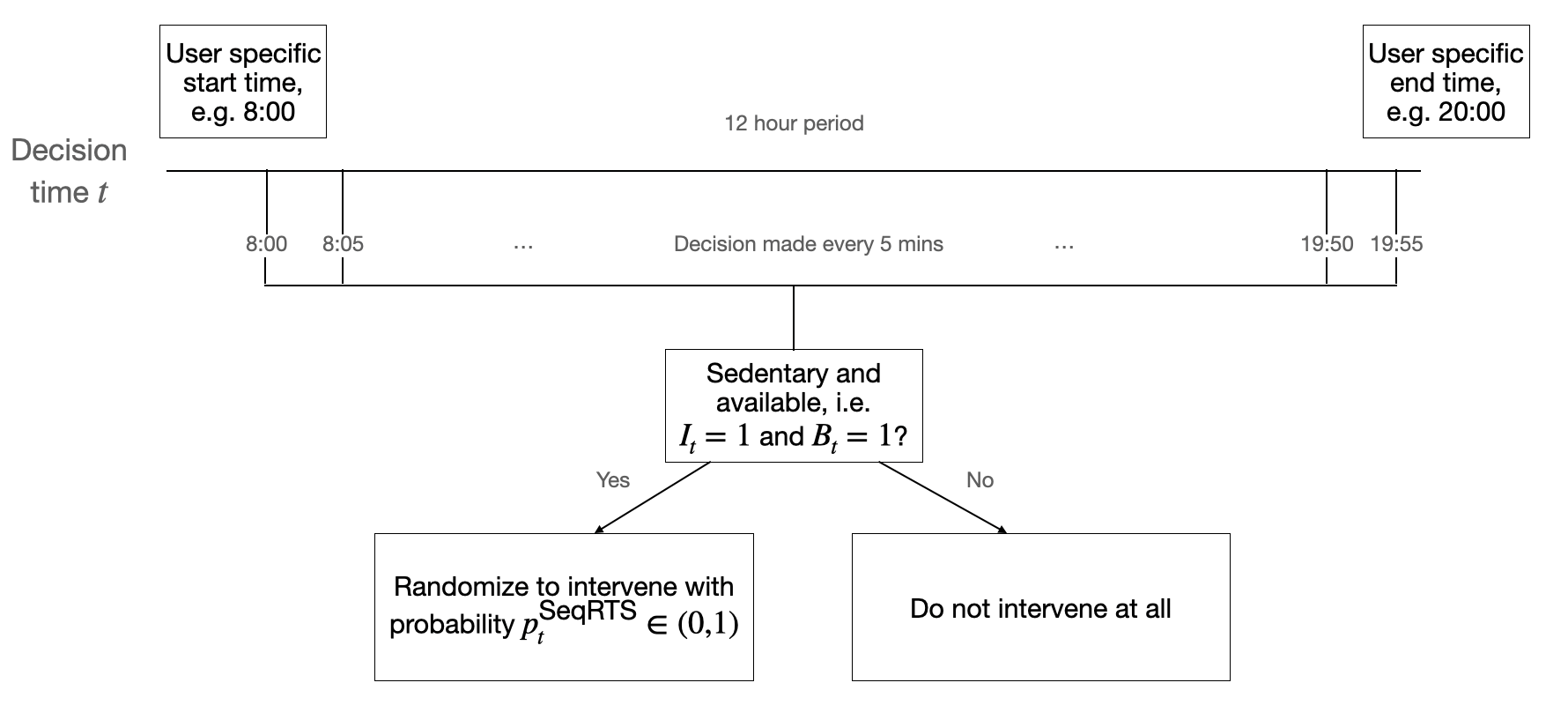}
    \caption{Decision times during a  typical participant's day in the HeartSteps V2V3 study.  $B_t=1$ if the participant is sedentary at decision time $t$; $I_t=1$ if the participant's decision time $t$ is available.}
    \label{fig:decision-day}
\end{figure}

See Figure \ref{fig:decision-day} for a typical participant's day.  The 82 participants were in the study (meaning from the first day they became eligible for randomization to the last day before they leave the study)  for a total duration of 11,316 participant days, leading to 1,629,504 (11,316 days $\times$ 144 decision times per day) decision times from June 2019 to  February 2021. Unfortunately, due to server-side errors from 15 February 2020 to 9 April 2020, step count data from the watch face was not transmitted to the cloud server; thus,  272,592 decision times (or 1,893 participant days) are not included. Before 15 February 2020, 82 participants contributed for 5,523 participant days. From April 9 to end of study, 36 participants contributed for 3,900 participant days. Together these two periods  resulted in 1,356,912  decision times (or 9,423 days). Of the 1,356,912 included decision times, participants were unavailable for treatment or not sedentary during 1,190,170 decision times (87.7\% of the time). Hence, data associated with  166,742 risk times (available and sedentary decision times) will be analyzed here.  Across the 166,742 risk times, 10,137 anti-sedentary messages were delivered.

\begin{figure}
      \centering
      \includegraphics[width=0.49\linewidth]{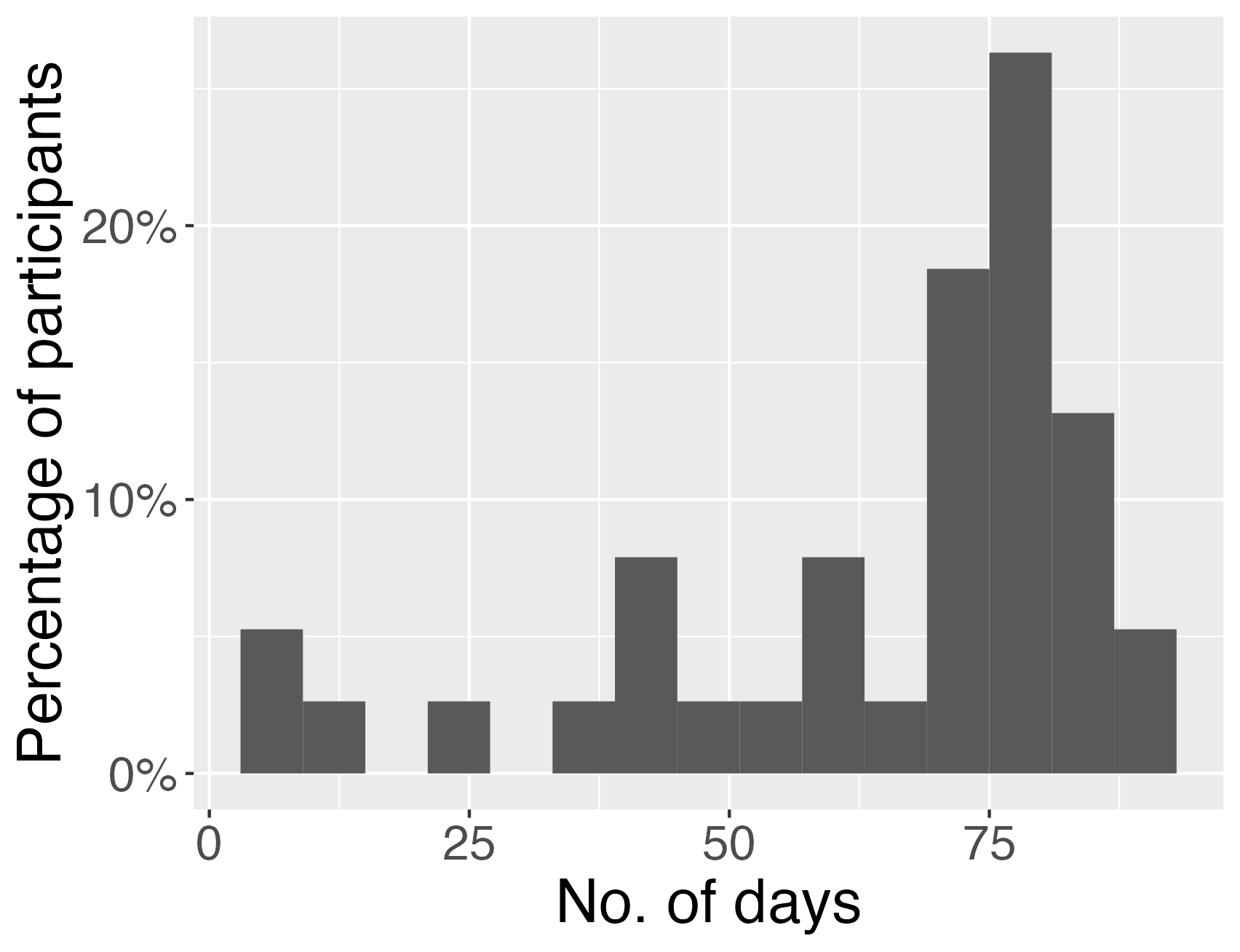}
      \includegraphics[width=0.49\linewidth]{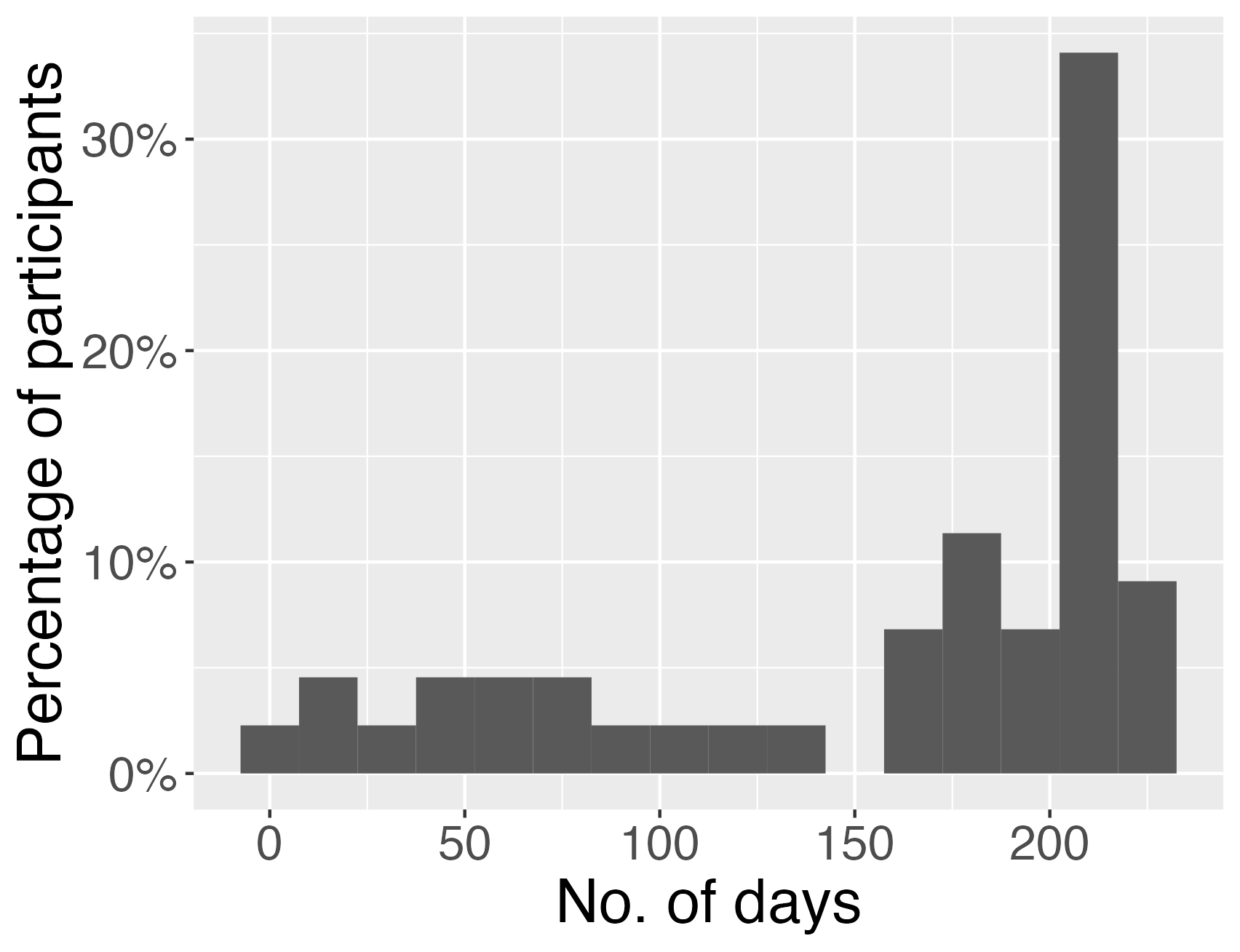}
      \includegraphics[width=0.49\linewidth]{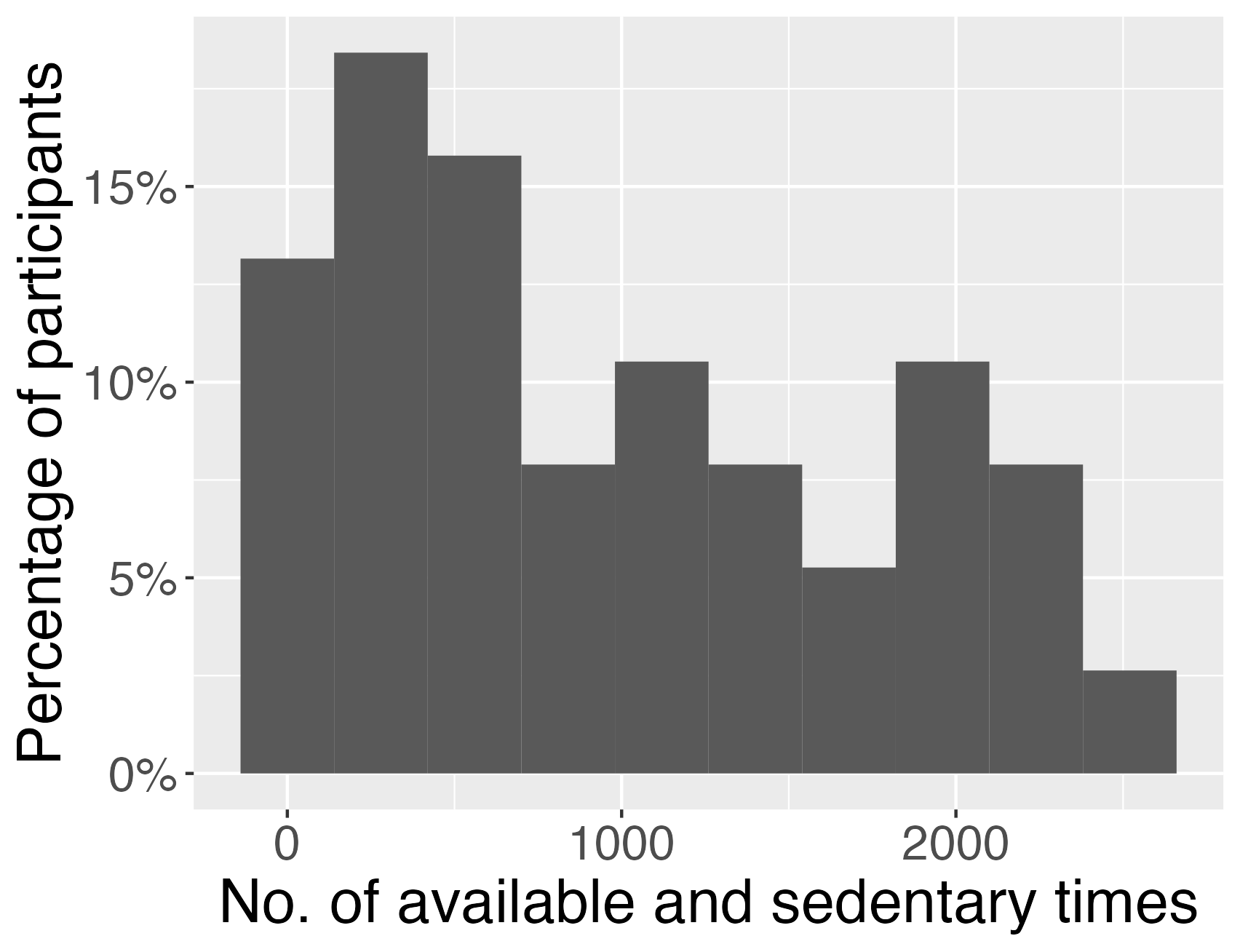}
      \includegraphics[width=0.49\linewidth]{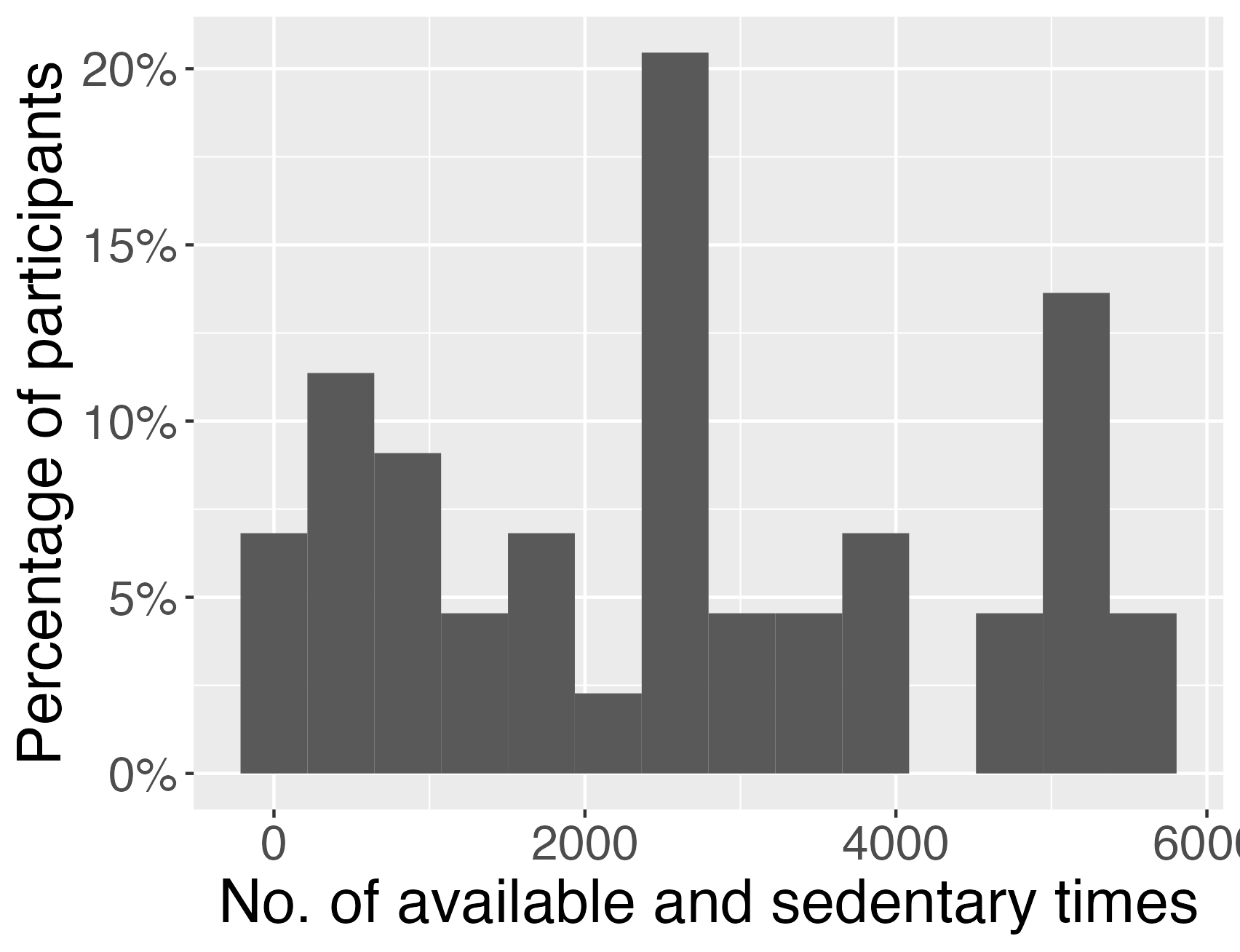}
      \caption{This figure consists of four histograms illustrating different aspects of participant activity in the study. From the top left to the bottom right: The first two histograms depict the distribution of the number of days each participant stayed in the study for V2 and V3, respectively. The third histogram showcases the distribution of the number of available and sedentary decision times per participant in V2, while the fourth histogram presents the distribution of the number of available and sedentary decision times per participant in V3.}
      \label{fig:hist-n-day-n-rand-per-user}
  \end{figure}

Figure \ref{fig:hist-n-day-n-rand-per-user} presents histograms of the number of days each participant stayed in the study for both V2 and V3, as well as histograms of the number of available and sedentary decision times per participant for V2 and V3. We notice that the modal number of days in the top row, which illustrates the histograms of the number of days per participant, does not align with the expected 90 days for V2 and 270 days for V3 participants. This discrepancy arises for two main reasons. Firstly, participants might leave the study early. Secondly, due to study consent constraints, data is retained for a maximum of 90 days for V2 and 270 days for V3 participants. This includes the baseline period when the intervention did not start. Another observation is that in the bottom row, the distribution of available and sedentary decision times is uneven across participants.

\section{The Sequential Risk Time Sampling (SeqRTS) Algorithm}
\label{sec:SeqRTS}

The ``Sequential Risk Time Sampling'' Algorithm (SeqRTS)  (\citeay{liao2018just},  \citeay{dempsey2020stratified}) was designed to satisfy two constraints, the average treatment constraint and the uniformity constraint. In HeartSteps V2V3, the average treatment constraint was an average of 1.5 anti-sedentary messages per day. This number was chosen by the scientific team to reduce burden on participant \citep{klasnja2008using, figueroa2021adaptive, liao2018just}. Each day is segmented into three equal four-hour blocks. SeqRTS works at the block level, and the goal of an average of 1.5 anti-sedentary messages per day translates into an average of 0.5 anti-sedentary messages per block. The uniformity constraint means the treatments should be delivered uniformly across all risk times. In this paper, we measure uniformity by checking how closely the randomization probabilities within each block resemble each other. See Section \ref{sec:measures} for our specific choice of the Mean Absolute Deviation.

We first introduce some notations before introducing SeqRTS formally. For $j=0,1,2,3$, let $\mathcal{T}_{i,j,d}$ denote the set of decision times for participant $i$, in block $j$ ($j=0$ means for the whole day), and on day $d$. In HeartSteps V2V3, $\mathcal{T}_{i,0,d} = \{1,2,...,144\}$, $\mathcal{T}_{i,1,d} = \{1,2,...,48\}$, $\mathcal{T}_{i,2,d} = \{49,50,...,96\}$, $\mathcal{T}_{i,2,d} = \{97,98,...,144\}$. Thus, $|\mathcal{T}_{i,0,d}| = 144$ and $|\mathcal{T}_{i,j,d}| = 48$ for $j=1,2,3$. We then use $T_0 = 144$ and $T = 48$ to denote number of decision times in a day or a block. Even though the beginning times in block 2 and 3 are not $1$, we use $t=1,.., T$ to count the times in all three blocks.  Denote $B_{i,j,d,t} \in\{0,1\}$ and  $I_{i,j,d,t}\in\{0,1\}$ as indicators of sedentary behavior and availability at time $t$, for participant  $i$, in block $j$, and on day $d$, where $B_{i,j,d,t}=1$ represents the participant is sedentary, $I_{i,j,d,t}=1$ represents the participant is available. A risk time is when participant is both sedentary and available, that is, $B_{i,j,d,t}=1$ and $I_{i,j,d,t}=1$. We denote the indicator of the risk time as $X_{i,j,d,t}$, and denote the total number of risk times in each block by $N_{i,j,d} = \sum_{t =1}^{T} X_{i,j,d,t}$. We denote time-varying covariates as $Z_{i,j,d,t}$. To simplify notations, we will often suppress the $i,j,d$ subscripts and refer to the indicators primarily as $B_t$, $I_t$, $X_t$, $Z_t$ unless the context necessitates the specification of the participant, block, or day. We also use bar notation to denote the vector of variables. For example, $\bar Z_t = \{Z_1, Z_2..., Z_t\}$.

We now use $N:= \sum_{t =1}^{T} X_{t}$ as the simplified notation for $N_{i,j,d}$, the number of risk times. When $N$ is known for a block at all risk times $\{t: X_{t} = 1\}$, the following sampling probability will achieve both constraints:
\begin{align}
    p_{t}^{Oracle} =\frac{0.5}{N}\label{eq:oracle}
\end{align}
This is because this probability is equal across all risk times, so the uniformity constraint is satisfied, and the probabilities sum up to 0.5, so the average treatment constraint is satisfied. We call Equation \ref{eq:oracle} the oracle probability. Note that although the oracle probability will achieve both constraints, it is impractical because the number of risk times, \(N\), is unknown at risk time \(t\). Specifically, \(N = \sum_{t' =1}^{T} X_{t'} = \sum_{t' =1}^{t-1} X_{t'} + \sum_{t' =t}^{T} X_{t'}\). The first term, \(\sum_{t' =1}^{t-1} X_{t'}\), is the number of risk times occurred before \(t\). It is known at time $t$. The second term, \(\sum_{t' =t}^{T} X_{t'}\), is the number of remaining risk times after time \(t\), which we denote as \(g_t\). \(g_t\) is unknown at time $t$, which makes \(N\) unknown.

If \( g_t \) is known, an alternative strategy that would achieve both constraints is as follows: We first refer to the average treatment allowed in the block as the ``budget," which we represent by \(N_0\) (\( N_0 = 0.5 \) in HSV2V3). At each decision time \( t \), we denote the remaining ``budget" by \( N_0 - \sum_{s = 1}^{t-1} X_s \cdot p_s \), where \( p_s \) is the treatment probability at time \( s \). At time \( t \), if we divide the remaining ``budget" by the remaining risk times $g_t$, the probability will be equal to the oracle probability (Equation \ref{eq:oracle}). This is the strategy that SeqRTS takes. See \cite{liao2018just} for a more detailed walkthrough.

Practically, \( g_t \) is not known at time \( t \), so it needs to be estimated \( g_t \). We represent the estimator of \( g_t \) as \( \hat{g}(H_t) \), where \( H_t \) denotes the set of time-varying variables collected so far in the block, i.e., \( H_t = \{\bar{Z}_t, \bar{A}_{t-1}, \bar{p}_{t-1}, \bar{I}_t, \bar{B}_t, \bar{X}_t\} \). To manage the estimation errors associated with $\hat{g_t}$, we make other quantities flexible. Although \( N_0 = 0.5 \) represents the concrete goal of the average number of treatments to be sent in a block, we make it a tuning parameter, \( \hat{N}_0 \), to cope with the uncertainty of \( \hat{g} \), ensuring that SeqRTS functions effectively. 
At decision time \( t \) in the \( j \)-th block on day \( d \), SeqRTS determines the probability of delivering treatment as follows:

\begin{equation}
    p_t^{SeqRTS}=\frac{\hat N_0 - \sum_{s = 1}^{t-1} X_s \cdot p^{SeqRTS}_s}{1 + \hat g(H_t)}\label{eq:p^{SeqRTS}}.
\end{equation}

In the parameter estimation stage, $\hat g$ was first estimated, then $\hat N_0$ was tuned given $\hat g$. The estimation and tuning for $\hat g$ and $\hat N_0$ were done using data from the prior study, Heartsteps V1 \citep[HSV1, ][]{klasnja2019efficacy}. $\hat g$ and $\hat N_0$ were never updated during HSV2V3.

The time-varying inputs used by $\hat g$ are a subset of $H_t$. These inputs include: (1) \textbf{Hour of the decision time:} $\text{Hour}_{t} := \lfloor t /12 \rfloor$; (2) \textbf{Risk ``run length":} $R_t$ represents the full length of time up until $t$ that a participant is at risk. $R_t$ can be defined recursively through the formula below:
\begin{align*}
R_t :=
  \begin{cases} 
   0 & \text{if } X_t = 0, \\
   R_{t-1} + 1 & \text{if } X_t = 1,
  \end{cases}
\end{align*}
where  $R_1$ is initialized to 0 to accommodate the beginning of the block. In Appendix \ref{appdix:hat-g}, we provide the exact form of $\hat g(H_t)$ used in HSV2V3.

$\hat N_0$ was tuned by selecting a value of $\hat N_0$ that satisfies two design constraints as closely as possible given a fixed $\hat g$. The best \((\hat{N}_0, \hat{g})\) pair \footnote{the pair under which simulated data based on HSV1 achieved an average treatment of 0.5 per block and minimized KL divergence between the treatment probabilities and the oracle probabilities. See Section~\ref{sec:measures} for a formal definition. } was selected, resulting in \(\hat{N}_0 = 1.8\). One reason to set a $\hat N_0$ that is much greater than $0.5$ is that $\hat g$ was generally too high, resulting in a higher denominator in Equation \ref{eq:p^{SeqRTS}} \footnote{See Appendix \ref{sec:quality-hatgt} for details}. Consequently, those blocks would generate fewer treatments than the target number of 0.5 if $\hat N_0$ were set to a lower value.  Lastly the probability, $p_t^{SeqRTS}$ is clipped to belong to the interval $[0.005, 0.2]$ to ensure the output value remains within a reasonable range and to avoid extreme probabilities that might lead to under- or over- treatment. 
The algorithm's developer chose 0.2 as an upper bound  based on an analysis of risk times in HSV1 data. 
Appendix \ref{appdix:hat-g} provides details for the choice of the choice.

Since \(\hat{N}_0\) and \(\hat{g}_t\) were overestimated, we are uncertain about the algorithm's performance in practice. \cite{liao2018just} suggests it should perform well on average across participants. In the next section, we will evaluate whether this expectation holds true.

\section{Evaluation of Algorithm's Performance on the Core Design Constraints}
\label{sec:evaluation}
In this section, we evaluate of SeqRTS's performance on the two core design constraints in HSV2V3, the average treatment constraint and the uniformity constraint. We first discuss the choice of data to use in evaluating the algorithm’s performance.  We then define measures of average treatment and uniformity. Finally, we evaluate the SeqRTS algorithm's ability to achieve an average of 1.5 treatments per day and its performance on maintaining uniformity across participants.

\subsection{Selection of Blocks and Days}
\label{sec:data-selection-avg-trt}

Before conducting our evaluation, we select blocks and days for each participant to accurately represent the algorithm's performance. Below, we mainly describe the procedure for selecting entire days, but the same criteria apply to the selection of blocks.

There are decision times, and indeed, entire days when participants are unavailable (see Section \ref{sec:HSV2V3-trial} for criteria for a participant's unavailability). Denote the availability indicator of a study day $d$ as \( I_d = (\sum_{t=1}^{T_0} I_{d,t} \geq 1) \), where there is at least one available decision time (with some abuse of notation, as we use \( I \) to represent both the availability of a decision time \( t \) and the availability of a study day \( d \)). For a participant, days we choose for evaluation are days that the participant is available: ${\cal D} = \{d: I_d = 1\}$. This selection of days results in 6,327 participant days out of the total 9,423 eligible participant days.

Days when a participant is not available are those in which \(I_d = 0\). They are excluded from our analysis because there is no data for the algorithm's performance on these days (since it could not perform). This means \( I_d \) an indicator of missing data. Inference based on the observed data alone might be biased if data is missing not at random \citep[MNAR; ][]{little2019statistical}. Here we argue that our selection of days does not suffer from this issue.

Recall the unavailability reasons described in Section \ref{sec:HSV2V3-trial}. When the participant is not available on a day, it means the participant was physically active, did not wear the watch, or that the watch face was not connected to the cloud server. We are not interested in days when a participant is physically active (e.g., when they go on a hike) because it makes no scientific sense to provide any anti-sedentary message when someone is active.
We are also not concerned with days when a participant did not wear the watch (e.g., due to busy schedules, travel, or simply forgetting), as there is no opportunity to assess sedentary behavior on those days. It is important to note that days when a participant does not wear the watch may be influenced by an unobserved variable \(U\) (e.g., mental tiredness). For example, if a participant is mentally tired, they may choose not to wear the watch and may also exhibit more sedentary behavior. On such days, the algorithm might not perform well due to \(\hat{g}_t\) being an inaccurate prediction.

As a result, the algorithm could appear to perform well by omitting days when a participant is mentally tired and not wearing the watch. We acknowledge this limitation, but note that our analysis of the HSV2V3 study is representative of Kaiser Permanente patients newly diagnosed with stage 1 hypertension. The algorithm’s performance is evaluated marginally across variables like \(U\), and therefore reflects the general behavioral patterns of the study population, including variability in watch-wearing behavior.

The days we are interested in are those when participants wear the watch and are not active all day. Missingness of the data on these days means the watch is not connected to the cloud server, which is caused by technical malfunctions of wearable devices, such as syncing issues, software glitches, i.e., technical reasons instead of user’s choices. Specifically, these reasons are not related to unmeasured factors like a participant’s mental state or unmeasured sedentary status on that day. Hence, we argue this missingness is at random (MAR), as it is either unrelated to the study variables or can be explained by observed data.

We apply the same logic to our choice of blocks. The number of days for which we have data from  blocks 1, 2, and 3 are 5,575, 5,473, and 4,994 respectively. Note that the number of included days for analysis of each block is smaller than the total of 6,327 participant days selected for analysis; if a participant day has an available decision time in block 1 but none in blocks 2 and 3, this day is included in the evaluation of block 1 and that of the day, but not in the evaluation of blocks 2 and 3.\footnote{See Section \ref{sec:daily-avg-less-block-avg} for a detailed illustration} We denote ${\cal D}_{j}$  as the set of blocks $j$ in which a  participant has an available decision time for $j=1,2,3$. When $j=0$ (whole day), we denote ${\cal D}_{0} = {\cal D}$

During the data analysis, we discovered a coding issue: for some decision times (most of them in block 3), a failure in correctly calculating \(\text{Hour}_t\), the hour of the decision time \(t\) (due to a coding issue in timezone conversion), caused \(\hat g(H_t)\) to be calculated incorrectly. As a result, at these decision times, the algorithm only produced a randomization probability of 0.005, the lower bound, which is very different from the correctly calculated randomization probabilities. Including too many of these decision times would improperly represent the algorithm’s performance on meeting the two constraints. We exclude 5.4\% of the blocks where there were many decision times at which \(\hat{g}_t\) was incorrectly calculated. The detailed exclusion criterion is presented in Table 4 in Appendix A.2, where this issue is described in more detail.

\subsection{Measures of number of treatments  and uniformity}
\label{sec:measures}

On day $d$, at each decision time $t$, the SeqRTS algorithm calculates a randomisation probability $p_{d,t}$ as detailed in Equation \ref{eq:p^{SeqRTS}}. An action $A_{d,t}$ is sampled from $p_{d,t}$ through $A_{d,t} \sim \text{Bern}(p_{d,t})$. The total number of treatments on day $d$  is calculated as $Y_d = \sum_{t=1}^{T_0}  A_{d,t}$. This $Y_d$ is used to evaluate the algorithm's performance on achieving 1.5 anti-sedentary messages per day.  Note that we are implicitly summing over risk times, since $p_{d,t}  > 0$ only at risk times.

In this paper, ``uniformity'' refers to how evenly treatments are distributed across all risk times. A sampling algorithm that achieves uniformity should generate treatment probabilities that are as similar as possible. Therefore, we measure uniformity using treatment probabilities. An ideal uniformity metric is zero when all probabilities are equal and increases as differences among probabilities grow. One such metric is the mean absolute difference (MAD), which measures the average absolute difference between probabilities at risk times and the mean probabilities on a day $d$:

\begin{equation}
     V_{d} =  \dfrac{\sum_{t: X_{d,t}=1} |p_{d,t} - \frac{1}{ N_{d}} \sum_{t': X_{d,t'}=1}  p_{d,t'}| }{ N_{d}}
    \label{eq:MAD}
\end{equation}

A smaller value of MAD represents more uniform probabilities. A perfect uniformity, i.e., when probabilities are equal on day $d$, means $V_{d} = 0$.

In past work, \cite{liao2018just} proposed evaluating uniformity using Kullback-Leibler (KL) divergence between SeqRTS probabilities and the oracle probabilities (Equation \ref{eq:oracle}) at each risk time. The KL divergence on a day \(d\) is an average of the KL divergence at all risk times in the day:
\begin{equation}
V_d^{KL} = \dfrac{\sum_{t: X_{d,t}=1} KL(p_{d,t},p_{d,t}^{Oracle}) }{ N_{d}}
\end{equation}

where $KL(p,q) = p \log\left(\frac{p}{q}\right) + (1-p) \log\left(\frac{1-p}{1-q}\right)$, and $ p_t^{Oracle} = 0.5 / N_{d} $.  They use \( V_d^{KL} = 0 \) to represent uniformity. However, \( V_d^{KL} \neq 0 \) even when uniformity is achieved. A toy example illustrates this phenomenon: Consider a block with two risk times. The oracle probabilities are \( 0.5 / 2 = 0.25 \) for both risk times. Assume the SeqRTS probabilities are  both \( 0.1 \) at the two risk times. The uniformity is achieved in this block since these two probabilities are equal, but they fail to meet the average treatment constraint, as their sum is \( 0.2 \), differing from the block's average treatment constraint of \( 0.5 \). A calculation shows that the MAD, \(V_d\) equals zero, but the KL divergence \(V_d^{KL} = 0.031\), is greater than zero. This demonstrates that KL divergence is affected by whether the average treatment constraint is met, in addition to the uniformity constraint. We discuss this observation in more detail in Appendix \ref{sec:discuss-no-KL}.

\subsection{Evaluate the Algorithm's Performance on the Two Constraints Over All Participants}

\subsubsection{Objectives, Assumptions and the Method}
The algorithm was designed to achieve, ideally, on average, 1.5 treatments per day with uniformity for each participant. This is a difficult goal to meet because the SeqRTS algorithm implemented in HSV2V3 applied a single set of \((\hat{g}, \hat{N}_0)\) to all participants and was never updated during the study, making it likely that the algorithm performs differently across participants. This is supported by the data.

\begin{figure}
	\centering
	\includegraphics[width=0.95\linewidth]{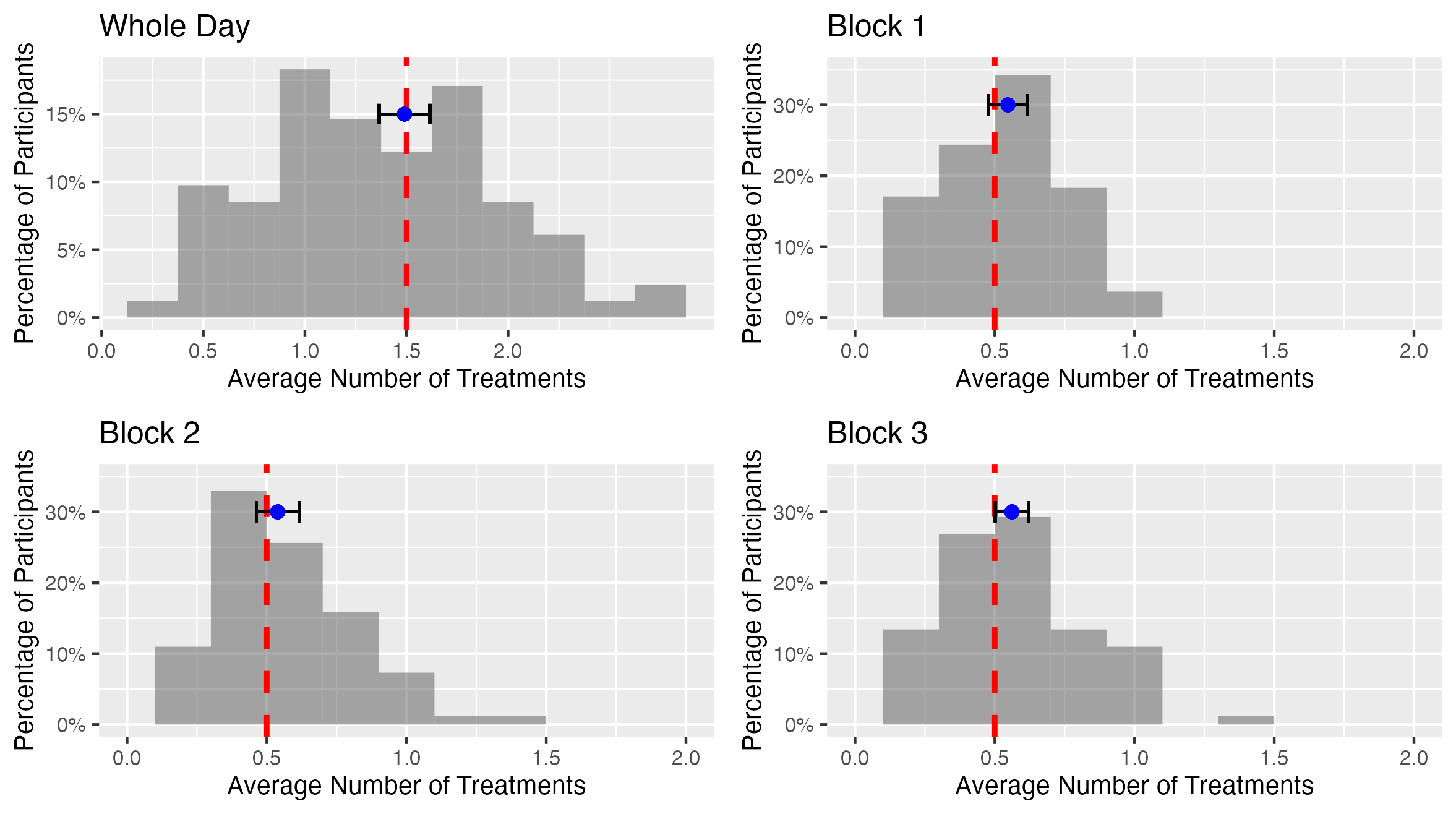}
	\caption{Histogram of the average number of treatments per participant for each block and the whole day for all 82 participants. Vertical lines indicate the block and daily goals (0.5 for each block and 1.5 for the whole day). The error bars represent the point estimates and 95\% confidence intervals of the average number of treatments across all participants. }
	\label{fig:avg-trt-by-blk}
\end{figure}

\begin{figure}
	\centering
	\includegraphics[width=0.95\linewidth]{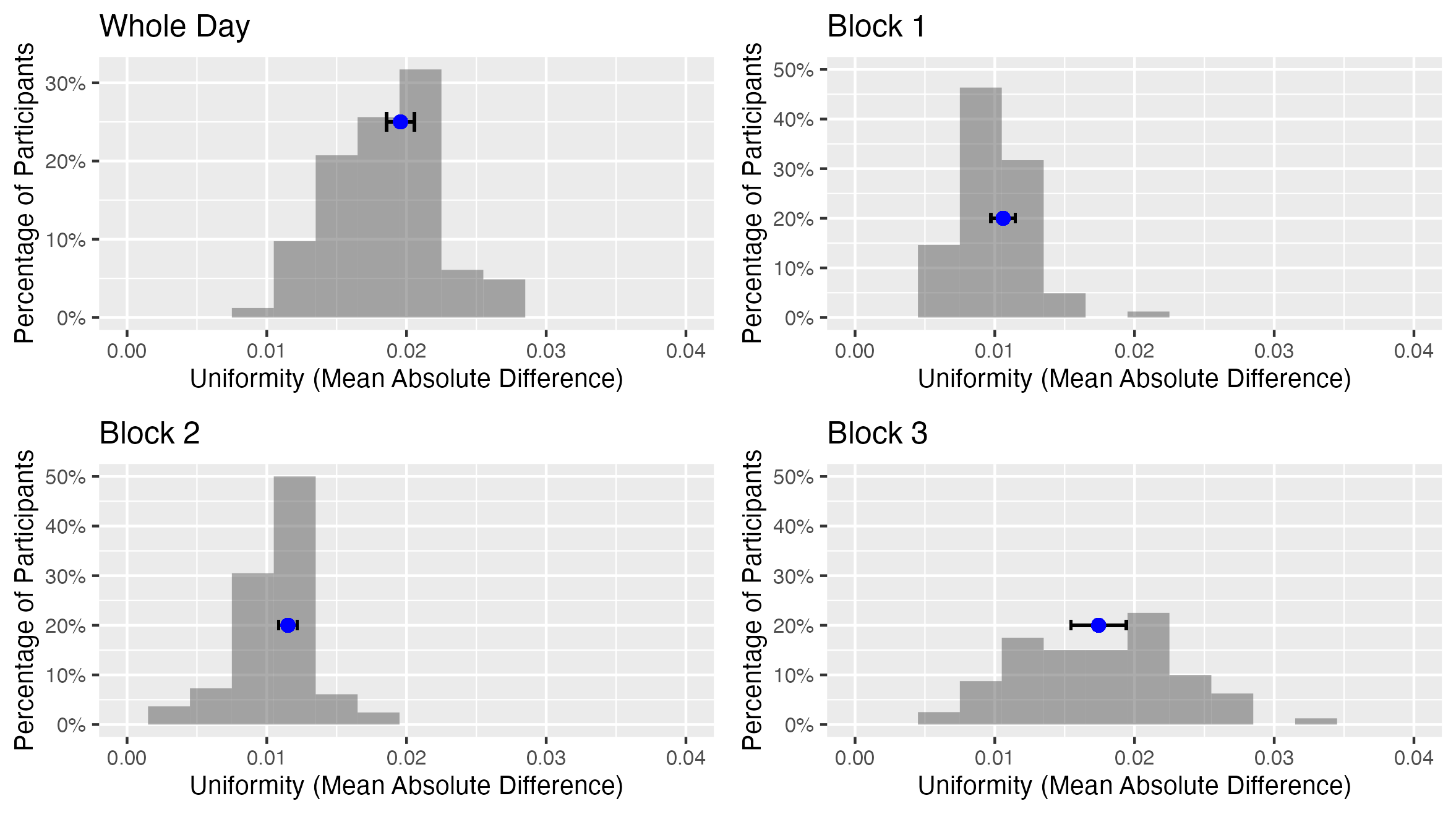}
	\caption{Histogram of participant average uniformity (mean absolute difference, MAD) for each block and the entire day for all 82 participants. The error bars represent the point estimates and 95\% confidence intervals of the MAD across all participants.}
	\label{fig:uniformity-by-blk}
\end{figure}

In Figure \ref{fig:avg-trt-by-blk}, we plot histograms of \(\bar{Y}_{i,j} = \frac{1}{|\mathcal{D}_j|}\sum_{d\in\mathcal{D}_j} Y_{i,j,d} \), the average number of treatments sent on the available days and blocks \(j = 0, 1, 2, 3\) over all participants \(i = 1, 2, \ldots, 82\). We also plot histograms of \(\bar{V}_{i,j} = \frac{1}{|\mathcal{D}_j|}\sum_{d\in\mathcal{D}_j} V_{i,j,d} \), the average MAD for all three blocks and the whole day, in Figure \ref{fig:uniformity-by-blk}. These histograms show that the algorithm's performance varies among participants, indicating that the two constraints were not met for each participant. 

A useful goal would be to assess the algorithm's performance in achieving these two constraints when averaged over all participants. This goal is important because if the algorithm achieves it, it demonstrates significant potential. In the rest of this section, we will investigate this using a statistical model.

To assess the goal, we perform inference on average number of treatments per day and average MAD on the whole population. These averages are over days when participants wear the watch and are not active all day (see Section \ref{sec:data-selection-avg-trt} for a discussion). Note that the number of treatments and the MAD for each participant \(i\), \(Y_{i,d}\) and $V_{i,d}$,  are outcomes that vary over days \(d\). We fit generalized estimating equations \citep[GEE; ][]{liang1986longitudinal, diggle2002analysis} to account for the longitudinal nature of the data.

We assume the population mean is constant throughout the study. An exploratory analysis shows that the number of treatments and MAD, when averaged over participants, do not vary across days. A test of significance on the study day variable using the same regression model as introduced later in Equation~\ref{eq:GEE-avg-trt}, but with the study day term added, further supports this assumption. We provide detailed analyses in Appendix \ref{sec:no-day-trend}. Additionally, we assume the treatment on the previous day does not affect sedentariness on the present day. This is supported by a data analysis in Appendix \ref{sec:CEE-treatment-sedentariness}, where we showed that there is insufficient evidence to indicate that anti-sedentary messages sent on one day have a causal effect on sedentariness the next day, when controlling for past sedentary behavior.

We now describe the inference procedure for the average treatment. A similar procedure is applied on the inference for the uniformity. We use the log mean model as our mean model to account for the count nature of $Y_{i,d}$. The model is presented in Equation \ref{eq:GEE-avg-trt}. Here, \(\mu_0\) represents the marginal mean, which is the average number of treatments sent across all participants. \(\eta\) is an over-dispersion parameter to account for the possibility that the data do not exactly follow a Poisson distribution (note that log mean model implies a Poisson regression). For simplicity, when presenting the results, we report only the standard error \(\sqrt{\hat \eta \hat{E}[Y_{i,d}] }\) and not \(\hat \eta\) (where the hat notation represents estimators).

\begin{equation}
\label{eq:GEE-avg-trt}
\begin{split}
    &  \log E[Y_{i,d}] = \log \mu_0  \\ 
   & \text{Var}(Y_{i,d}) =  \eta E[Y_{i,d}]
\end{split}
\end{equation}
We fit the model using both an independent working correlation model and a first-order autoregressive model (AR-1) model. The AR-1 model accounts for the correlation between outcomes on different days, where outcomes closer in time are more correlated than those further apart. It is particularly useful for longitudinal data, where outcomes (number of treatments and MAD) may be correlated within each participant and across days. The AR-1 assumption does not need to be correct, but the estimation will be more efficient when it is correct.

For MAD \(V_{i,d}\), we fit the GEE using an identity link to estimate the mean MAD across all participants. This is because uniformity is a continuous variable, and exploratory analysis shows that the histogram of MAD for each participant across all days is approximately bell-shaped.

Recall from Section~\ref{sec:SeqRTS} that the algorithm operates at the block level. Besides the daily goals, we also analyze whether the algorithm achieves the block goals. The analyses for each block and the whole day were conducted separately. We conclude the section by noting that the validity of the GEE results relies on the assumption that data is missing completely at random (MCAR). Appendix \ref{sec:MAR-vs-MCAR} shows that this is a plausible assumption.

\subsubsection{Results}

We now present the evaluation results for average treatment and uniformity.  The first two columns in Table \ref{tab:gee-marginal} present the estimation results on average number of treatments across users. We make two observations. First, the 95\% confidence interval covers the goals (0.5 or 1.5), except for block 3, where the average is slightly greater than the goal of 0.5\footnote{Note that the block averages do not sum up to the daily average. This discrepancy arises from the possibility that a participant is available during one block but not the others on a specific day. Consequently, the daily average, which accounts for a greater number of days, is lower than the sum of the block averages. A detailed illustrative example can be found in Appendix \ref{sec:daily-avg-less-block-avg}.}. Second, both AR-1 and independence models produce similar estimation results. The AR-1 correlation coefficient is significant (p-values for all blocks, which are omitted from the table, are less than 0.05), indicating that outcomes in consecutive days are likely correlated. Given this plausible correlation between consecutive days, we present the results using the AR-1 model in the following sections of the paper.

It is a significant result for SeqRTS to achieve the average treatment constraint across all users because the third unavailability criterion—unavailability for one hour after each treatment—created additional difficulty. The difficulty is created because the time when a treatment is delivered is an unknown event, even to the oracle, who knows the number of risk times. When a treatment is delivered, the number of risk times could be reduced, and hence the total number of treatments delivered in a block could be lower than 0.5 for the oracle. Indeed, in Appendix \ref{sec:robustness-performance}, we use a simulation to show that on all available blocks, the oracle algorithm delivers an average number of treatments lower than the goal of 0.5 across all participants. This demonstrates that SeqRTS outperforms the oracle approach in handling real-world complexities.

\begin{table}[ht]
\centering
\caption{Average treatment and uniformity for each block and the entire day across all 82 participants, estimated by generalized estimating equations (GEE). Each cell contains the point estimate and the 95\% confidence interval (presented in the bracket). }
\label{tab:gee-marginal}
\begin{adjustbox}{width=\textwidth}
\begin{tabular}{llcccc}
\toprule
 &  & \multicolumn{2}{c}{Average Number of Treatments} & \multicolumn{2}{c}{Uniformity (MAD)} \\
\cmidrule(lr){3-4} \cmidrule(lr){5-6}
Block / Working Correlation &  & Independence & AR-1 & Independence & AR-1 \\
\midrule
Whole Day &  & 1.490 (1.365, 1.615) & 1.486 (1.365, 1.607) & 0.019 (0.018, 0.020) & 0.019 (0.018, 0.020) \\
Block 1 &  & 0.547 (0.477, 0.617) & 0.552 (0.483, 0.621) & 0.010 (0.009, 0.011) & 0.010 (0.009, 0.011) \\
Block 2 &  & 0.539 (0.462, 0.615) & 0.544 (0.469, 0.618) & 0.011 (0.010, 0.011) & 0.011 (0.010, 0.011) \\
Block 3 &  & 0.562 (0.501, 0.622) & 0.569 (0.511, 0.627) & 0.017 (0.016, 0.019) & 0.017 (0.016, 0.019) \\
\bottomrule
\end{tabular}
\end{adjustbox}
\end{table}

The last two columns in Table \ref{tab:gee-marginal} present the estimation results on average MAD across users. Blocks 1 and 2 have an average MAD of approximately 0.01, but block 3 exhibits a higher MAD of approximately 0.018. The larger MAD in block 3 is due to the presence of risk times with a lower bound probability of 0.005, which differs significantly from other probabilities. This is a result of the coding issue mentioned in Section~\ref{sec:data-selection-avg-trt} and detailed in Appendix \ref{sec:coding-issue}. As described in Section~\ref{sec:data-selection-avg-trt}, days with large proportions of coding errors were omitted, but days with mildly large proportions of coding errors were retained, primarily affecting block 3 but not blocks 1 and 2.

Unlike the average treatment constraint, where a target is established, determining whether MAD is low is relatively subjective. Note that \(\text{MAD} = 0.01\) for a block means that, on average, treatment probabilities within the block differ from their block average by only 0.01, which we interpret as good uniformity. \(\text{MAD} = 0.018\) also indicates good uniformity, especially considering that the probabilities sum up to 0.5 on average. As such, average MAD across all participants of 0.01 (for blocks 1 and 2) and 0.018 (for block 3) indicates good uniformity.

\section{Identify key covariates to improve the online algorithm's performance on meeting the core design constraints}
\label{sec:covars}
The previous section highlighted the algorithm’s good performance in meeting both the average treatment and uniformity constraints across all participants. However, the algorithm’s performance did not meet the ultimate objective of fulfilling these constraints for each individual participant: as Figure \ref{fig:avg-trt-by-blk} and \ref{fig:uniformity-by-blk} show, there are participants for whom the algorithm’s performance is suboptimal.

In this section, we discuss which baseline (Section \ref{sec:CART-bsl-covar}) or time-varying covariates (Section \ref{sec:time-varying-feature}) could be used to improve to algorithm's performance in meeting constraints for each individual participant. We will do so by figuring out which covariate correlates with the algorithm's performance metrics (average treatment and MAD) in HSV2V3. To figure out these variables, we first come up with candidate variables: for baseline variables, the scientific team suggested a list of candidates. For time-varying variables, we examine specific aspects of SeqRTS, namely the formula itself and factors that leads to variation of $\hat g_t$ that affects performance on achieving two constraints. After we have a list of candidate variables, statistical models are fit to both select baseline covariates from the candidate and identify the population for whom the current algorithm under-performed. For baseline variables, we use a Classification and Regression Trees \citep[CART; ][]{breiman2017classification}. For time-varying covariates, we use GEE \citep{liang1986longitudinal}  to  identify the blocks for which the current algorithm under-performed (Section \ref{sec:avg-trt-gee}, \ref{sec:uniformity-gee}). Last but not least, we make suggestions on directions for the algorithm's improvement after identifying the covariates.

\subsection{Using CART to Identify the Population for Whom the Current Algorithm Under-performed}
\label{sec:CART-bsl-covar}

\subsubsection{Describe Candidate Baseline Covariates}
\label{sec:candidate-covar}
The scientific team identified candidate variables thought to be correlated with sedentary behavior, and thus could be correlated with algorithm's performance. These variables include 1. \texttt{gender}: captures the respondents' self-identified gender based on the options provided in the questionnaire. The collected data for the ``gender" variable contains responses of ``female" or ``male." 2. \texttt{is retired}:  Whether a participant is retired. 3. \texttt{sedentary behavior}: How many hours a participant typically spends sitting on a weekday in the past 7 days. 4. \texttt{activity level}: median daily total number of steps during the baseline period.

To ensure interpretability of the fitted model, we categorize numeric values with meaningful cutoff. We categorize activity level into ``Low Activity", ``Moderate Activity", ``High Activity", and ``Very High Activity" categories, as delineated by \cite{tudor2004many}, using the cut-offs 5000, 7500, 9000. We also categorize sedentary behavior into  ``Low Sedentary", ``Moderate Sedentary", ``High Sedentary" using cutoffs of 4 and 8 sedentary hours in a day \citep{biswas2015sedentary, owen2010too}. Details on the rationale why and how we do this are discussed in Appendix \ref{sec:bsl-categorize}.  

There are 9 participants for whom sedentary behavior data in the baseline covariates are missing. To address this missingness, we applied three methods—K-Nearest Neighbor (KNN) imputation, median imputation, and omission of missing entries—resulting in the \texttt{df\_KNN}, \texttt{df\_median}, and \texttt{df\_drop} datasets, respectively. Details of how missingness was handled can be found in Appendix \ref{sec:bsl-missingness}.

\subsubsection{Using CART to select baseline covariates}
\label{sec:avg-trt-bsl}
The model we used to both select baseline covariates and identify the population for whom the current algorithm under-performed are the tree models: Classification and Regression Trees (CART) and random forests \citep{breiman2001random}. The tree models are selected in this analysis because they offer interpretable classification criteria. This is important for distinguishing participants for which the algorithm under-performed because each leaf node with be a sub-population with different performance. 

The analysis proceeds in two steps. First, we conduct a stability check using leave-one-out cross validation (LOOCV) to select hyper-parameters, in other words, select a model to use. Then, with the selected model, we fit to the data and interpret the result. We conduct analysis on average treatment constraint and uniformity constraint separately. Since the analysis procedures are the same, we mainly present the analysis procedure for average treatment constraint and just report the result for the uniformity constraint.

\textbf{First step: stability check}
We focus on the analysis of average treatment. For average treatment, we use $\bar Y_{i,j}$, average treatment for participant $i$ in block $j$ as the outcome, for all $j= 0,1,2,3$ To find an optimal model, we employ leave-one-out cross validation (LOOCV) to select the hyper-parameter. The rpart package in R \citep{therneau2015package} mitigates CART's over-fitting by utilizing the complexity parameter (cp). We vary the fitting of trees with cp values ranging from 0.01 to 0.07. We also fit random forests, which have fewer over-fitting concerns, using lengths between 3 and 7. The resulting nodes symbolize participant clusters that share similar treatment averages. 

In Table \ref{tab:LOOCV-avg-trt}, we report the LOOCV training and testing mean squared error (MSE) and the average split count for all combinations fitted on the three datasets, df\_KNN, df\_median, and df\_drop.  The table presents results for average treatment, with a focus on interpreting the model’s stability. For the CART model, cp values of 0.05 and 0.07 demonstrate greater stability, showing a narrower gap between training MSE and validation MSE, which suggests less overfitting compared to cp values of 0.01 or 0.03. Additionally, the average number of nodes is lower for cp values of 0.05 and 0.07, indicating a simpler model structure. These conclusions hold across all three datasets. 

For the random forest model, setting maxnodes to 3 achieves closer training and validation errors, comparable to the CART model with cp values of 0.05 and 0.07, especially for the df\_drop and df\_KNN datasets. However, the df\_median dataset shows a slight discrepancy, where the CART model with cp values of 0.05 and 0.07 has a slightly higher validation error than the random forest with maxnodes set to 3, possibly due to sub-optimal imputation of missing data in the df\_median dataset. In conclusion, we limited the number of nodes to around 3 to 4.

To ensure stable model performance, we selected a cp value of 0.07 for fitting the tree model, using only the df\_drop and df\_KNN datasets. These datasets exhibited a narrower gap between training MSE and validation MSE during the stability checks, making them ideal candidates for this phase. In contrast, the df\_median dataset was excluded due to concerns about potential inaccuracies from sub-optimal imputation of missing data, as indicated by the larger gap between training and validation MSE.

\begin{table}
    \centering
    \caption{Prediction results for on Leave-One-Out Cross-Validation (LOOCV) for datasets created from three imputation methods (df\_drop, df\_KNN, df\_median) over two interpretable models (CART and random forest) with various hyper-parameters. CART is evaluated with complexity parameters (cp) ranging from 0.01 to 0.07, and random forest with maxnodes ranging from 3 to 7. Panel A reports the training MSE, Panel B reports the validation MSE, and Panel C shows the mean number of nodes resulted in the models.}
    \label{tab:LOOCV-avg-trt}
\begin{tabular}{lccccccccc}
    \toprule
    & \multicolumn{4}{c}{CART (cp)} & \multicolumn{4}{c}{random forest (maxnodes)} \\
    \midrule
    & 0.01 & 0.03 & 0.05 & 0.07 & 3 & 5 & 7 \\
    \midrule
    \multicolumn{8}{l}{\textit{Panel A: Training MSE}}\\
    df\_drop   & 0.197 & 0.212 & 0.220 & 0.222 & 0.263 & 0.225 & 0.200 \\
    df\_KNN    & 0.183 & 0.198 & 0.214 & 0.215 & 0.259 & 0.226 & 0.204 \\
    df\_median & 0.201 & 0.218 & 0.224 & 0.226 & 0.260 & 0.227 & 0.206 \\
    \midrule
     \multicolumn{8}{l}{\textit{Panel B: Validation MSE}} \\
    df\_drop   & 0.300 & 0.312 & 0.300 & 0.299 & 0.302 & 0.282 & 0.276 \\
    df\_KNN    & 0.335 & 0.350 & 0.313 & 0.313 & 0.295 & 0.278 & 0.273 \\
    df\_median & 0.312 & 0.335 & 0.344 & 0.338 & 0.297 & 0.281 & 0.275 \\
    \midrule
     \multicolumn{8}{l}{\textit{Panel C: Mean number of nodes}} \\
    df\_drop   & 10.0 & 5.67 & 4.03 & 3.86 & 3.00 & 5.00 & 7.00 \\
    df\_KNN    & 11.1 & 7.02 & 3.98 & 3.88 & 3.00 & 5.00 & 7.00 \\
    df\_median & 10.3 & 4.90 & 3.90 & 3.63 & 3.00 & 5.00 & 7.00 \\
    \bottomrule
\end{tabular}
\end{table}

\begin{figure}
    \centering
    \includegraphics[width=0.49\linewidth]{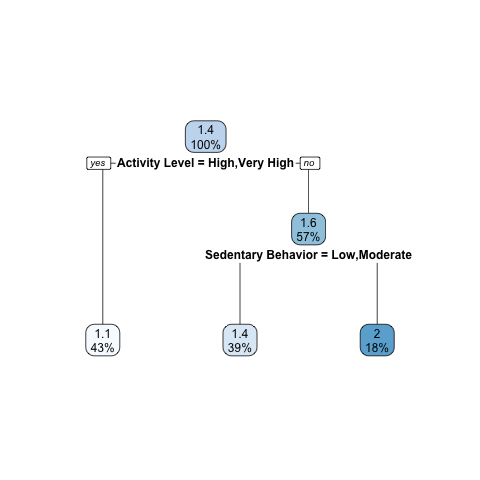}
    \includegraphics[width=0.49\linewidth]{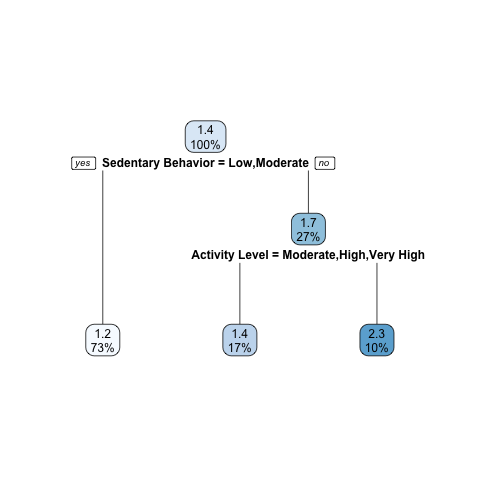}
    \caption{Decision tree for fitting df\_drop (tree 1, on the left) and df\_KNN (tree 2, on the right). Each node contains a number representing the average treatment for participants in the node, and the proportion of participants in the node. Both trees utilize sedentary behavior and activity level to make splits. Tree 1 was preferred due to its stability.}\label{fig:user_lvl_EDA_plots_tree_avg_trt}
\end{figure}

\textbf{Second step: Select meaningful sub-population where the algorithm's performance is different}
After selecting the model, we fit the tree model with a cp of 0.07 on both the df\_drop and df\_KNN datasets and interpret the results.
The decision tree is depicted in Figure \ref{fig:user_lvl_EDA_plots_tree_avg_trt}.  The outcomes of tree fitting on the df\_drop and df\_KNN datasets resulted in the formulation of two distinct trees: tree 1 and tree 2. To further assess stability, we randomly excluded 4 participants across 5 different random seeds and refit the data on both df\_drop and df\_KNN. The df\_drop dataset consistently yielded tree 1, while the df\_KNN dataset also produced tree 1 in 4 out of 5 instances, instead of tree 2. As a result, tree 1 was selected as the final model due to its demonstrated stability.

Tree 1 categorizes participants into three distinct groups, characterized by average treatment values of 1.1, 1.4, and 2. These groups can be intuitively labeled as under-treated, well-treated, and over-treated, respectively. The splitting variables are activity level and sedentary behavior. The first splitting criterion is based on activity level: if a participant has high or very high activity (daily step counts equal to or exceeding 7500), they are under-treated, receiving an average of 1.1 messages per day. This makes sense, as active participants are less available and thus have fewer opportunities to be randomized. The second criterion is based on sedentary behavior: if a participant does not have high or very high activity and is highly sedentary (sedentary hours equal to or exceeding 8 hours), they will be over-treated, receiving an average of 2 messages per day. This is logical, as being more sedentary provides more opportunities for randomization and, consequently, more treatments.

\subsubsection{Baseline Variables Are Not Correlated with Uniformity}
\label{sec:uniformity-bsl}
We repeat the same procedure with the uniformity data. The result indicates that none of the baseline variables are predictive of uniformity, demonstrating that the uniformity is not influenced by the baseline variables we considered.

\subsection{Using GEE to Identify Influential Time-Varying Covariates}
\label{sec:time-varying-feature}
In this section, we identify time-varying covariates that could be used to improve the algorithm’s performance in meeting constraints for each individual participant. We do so by figuring out which covariates correlate with average treatment and MAD on all 9,964 participant blocks. We use Generalized Estimating Equations (GEE) to conduct analysis. We present the analysis of the average treatment constraint and the uniformity constraint in Sections \ref{sec:avg-trt-gee}, \ref{sec:uniformity-gee} respectively.

\subsubsection{Identify Time-varying Covariates to Improve Average Treatment}
\label{sec:avg-trt-gee}
In this section, we explore potential time-varying covariates that may improve SeqRTS's performance in achieving an average treatment number of 0.5 per block for each participant. We first identify and construct time-varying covariates that are correlated with average treatment over blocks, then fit a GEE regression model to examine their relationship with the number of treatments sent in the block. Lastly, we analyze the regression results and make a recommendation on how to improve SeqRTS. We construct these time-varying covariates by examining the numerator and denominator of the SeqRTS formula (Eq. \ref{eq:p^{SeqRTS}}) separately.

First, we examine the numerator. The numerator, \(\hat{N}_0\), the approximate average treatment allowed in the block, is the same across all blocks. In Section 3, the explanation of how SeqRTS works shows that the sum of treatment probabilities in a block depends on how much of \(\hat{N}_0\) is spent. For blocks with more risk times, SeqRTS has more opportunities to spend \(\hat{N}_0\). Conversely, fewer risk times offer fewer opportunities to spend \(\hat{N}_0\). This suggests that ``riskness" is a relevant time-varying covariate to use in the regression. We refer to ``riskness" as ``sedentariness" in the remaining part of the section, since sedentariness is a more descriptive term for a participant's behavior. Formally, we define mean sedentariness\(\text{MeanSed} = \sum_{t=1}^T \tilde{X}_t\) (recall \(T = 48\) represents the number of decision times in a block). 
We hypothesize that larger MeanSed is positively correlated with the average treatment.

\begin{figure}
    \centering
    \includegraphics[width=\linewidth]{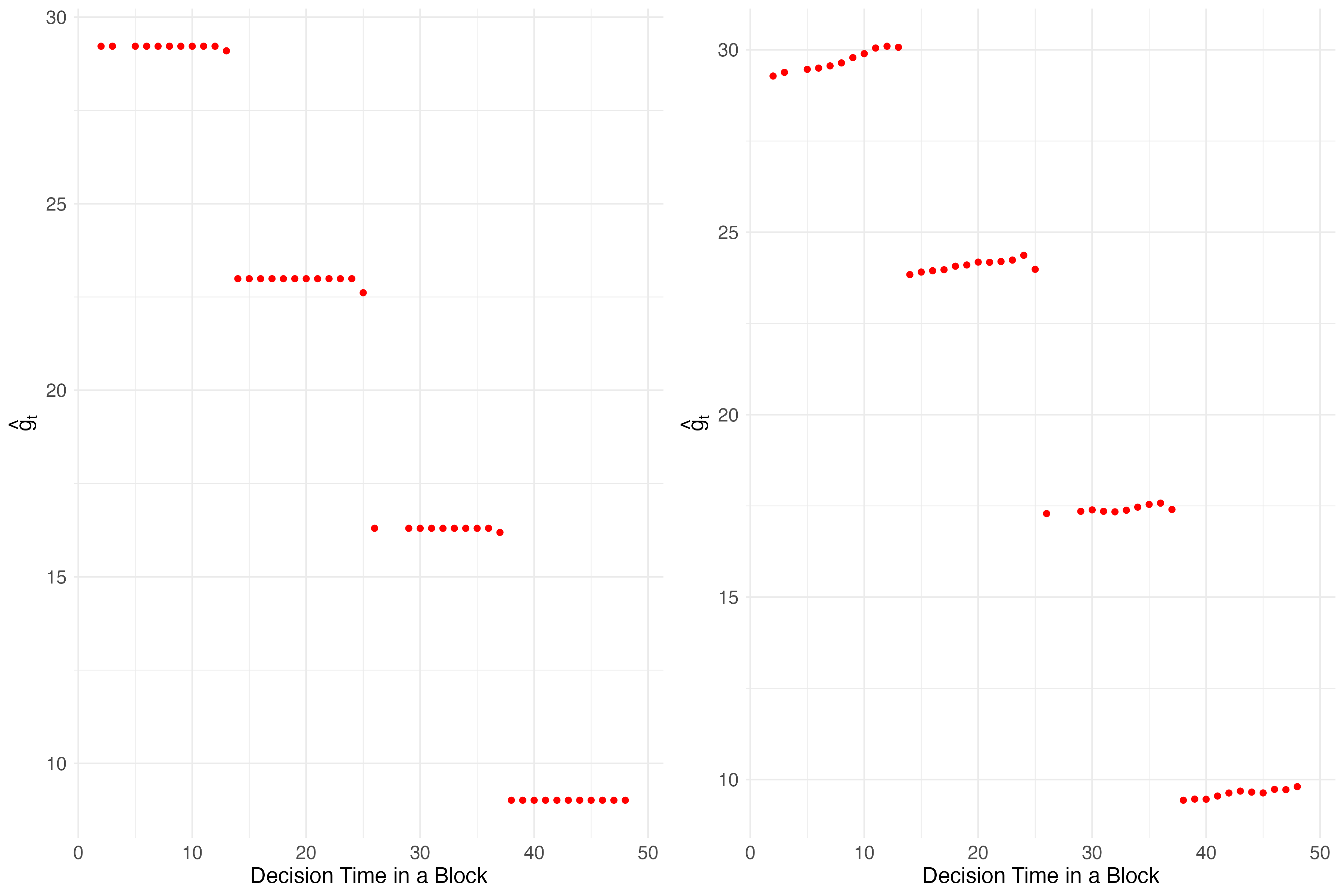}
    \caption{Plot of \(\hat{g}_t\), the estimated remaining risk times, over \(t\) in a block. We plot \(\hat{g}_t\) versus decision time in a block, varying two sets of values for \(R_t\), the full length of time up until $t$ that a participant is at risk, or the risk ``run length", one of two variables that \(\hat{g}_t\) uses. The two sets of values are the lower and upper bounds of \(R_t\): a) \(R_t = 1\) for each decision time (the left plot), i.e., assuming the participant has not been sedentary up until $t$, and b) \(R_t = t\), i.e., assuming the participant has been sedentary all the time prior to time $t$ (the right plot). Each 12 decision times is an hour. These two plots show that \(\hat{g}_t\) changes the most after 12 decision times, i.e., when $t$ goes to a new hour. Specifically, \(\hat{g}_t\) is smaller when \(t\) falls in the later hours of the block.}
    \label{fig:hat_g_t}
\end{figure}

Second, we examine the denominator \(\hat{g}_t\), the estimated remaining risk times after time \(t\). Our construction of \(\hat{g}_t\), as illustrated in Figure \ref{fig:hat_g_t}, shows that \(\hat{g}_t\) is smaller when \(t\) falls in the later hours of the block. 
The reason \(\hat{g}_t\) is smaller near the end of the block is that there are fewer decision times near the end. When \(\hat{g}_t\), the denominator is small, the treatment probability at time \(t\) is large. This suggests another time-varying covariate: end block sedentary proportion, or \(\text{EndProp}\): Let $N$ be the number of risk times in block. Then EndProp is the fraction of these risk times appearing at one of the last $N$ decision times in a block. Mathematically, \(\text{EndProp} = \frac{\sum_{t= T-N+1}^T \tilde{X}_t}{N}\). \(\text{EndProp}\) quantifies how much of the risk times appear at the end segment of a block. We hypothesize that the larger EndProp is, the larger the number of treatments is on average.

Before conducting the GEE regression, we took a few steps to ensure that there is no endogeneity, where endogeneity refers to the correlation between the predictors and the errors that can undermine the validity of the GEE analysis by giving biased and inconsistent estimates \citep{diggle2002analysis}. First, note that the MeanSed and EndProp are functions of the risk time indicator \(X_t\). They are is susceptible to endogeneity: due to the one-hour unavailability following a treatment (the third cause of unavailability in Section~\ref{sec:HSV2V3-trial}), blocks with at least one treatment may have a different risk time distribution compared to blocks without any treatment, even if they have the same degree of sedentariness. As a result, we impute an availability indicator based on all other reasons for unavailability, excluding the third cause, by setting the user to be available for all decision times within one hour following a treatment \footnote{Details can be found in Appendix \ref{appendix:imputation-I_t}.}. The corresponding risk indicator is denoted as $\tilde X_t$. Second, we use the prior 5-day average of the time-varying covariates instead of their same-day values. This is to remove reverse causality and ensure that the number of treatments does not influence the two time-varying covariates, thus ensuring no endogeneity in the regression. Lastly, we assume  the treatment has no effect on sedentary behavior when controlling for past sedentary behavior. This assumption is supported by a data analysis in Appendix \ref{sec:CEE-treatment-sedentariness}.

We now use GEE regression to examine the relationship between MeanSed, EndProp and the number of treatments in a block, $Y$. An exploratory data analysis that considers all first and second-order terms of $\text{MeanSed}$ and $\text{EndProp}$ reveals that $\text{MeanSed}$, $\text{MeanSed}^2$, and $\text{EndProp}$ are correlated with $Y$. We then fit a GEE with a log-linear mean model (to accommodate the count nature of the response variable $Y$) to examine the significance of $\text{MeanSed}$, $\text{MeanSed}^2$, and $\text{EndProp}$, and to quantify their relationships with $Y$. We implemented various correlation structures, including 'independent', 'AR-1', and 'exchangeable'. The results were consistent across these different specifications. We choose 'AR-1' because it is a plausible correlation structure for the outcome, as discussed earlier in Section 4.3. To check residuals, we set aside  the last 20\% of days as testing data for each participant, then fitted a locally estimated scatterplot smoothing (LOESS) curve between the fitted values and the Pearson residuals \citep{diggle2002analysis}. The residuals were closely aligned with zero on the residual plots for both the training and testing datasets. The residual plots of the GEE model shows no evidence of a relationship between the fitted values and the residuals.

\begin{figure}
    \centering\includegraphics[width=0.4\linewidth]{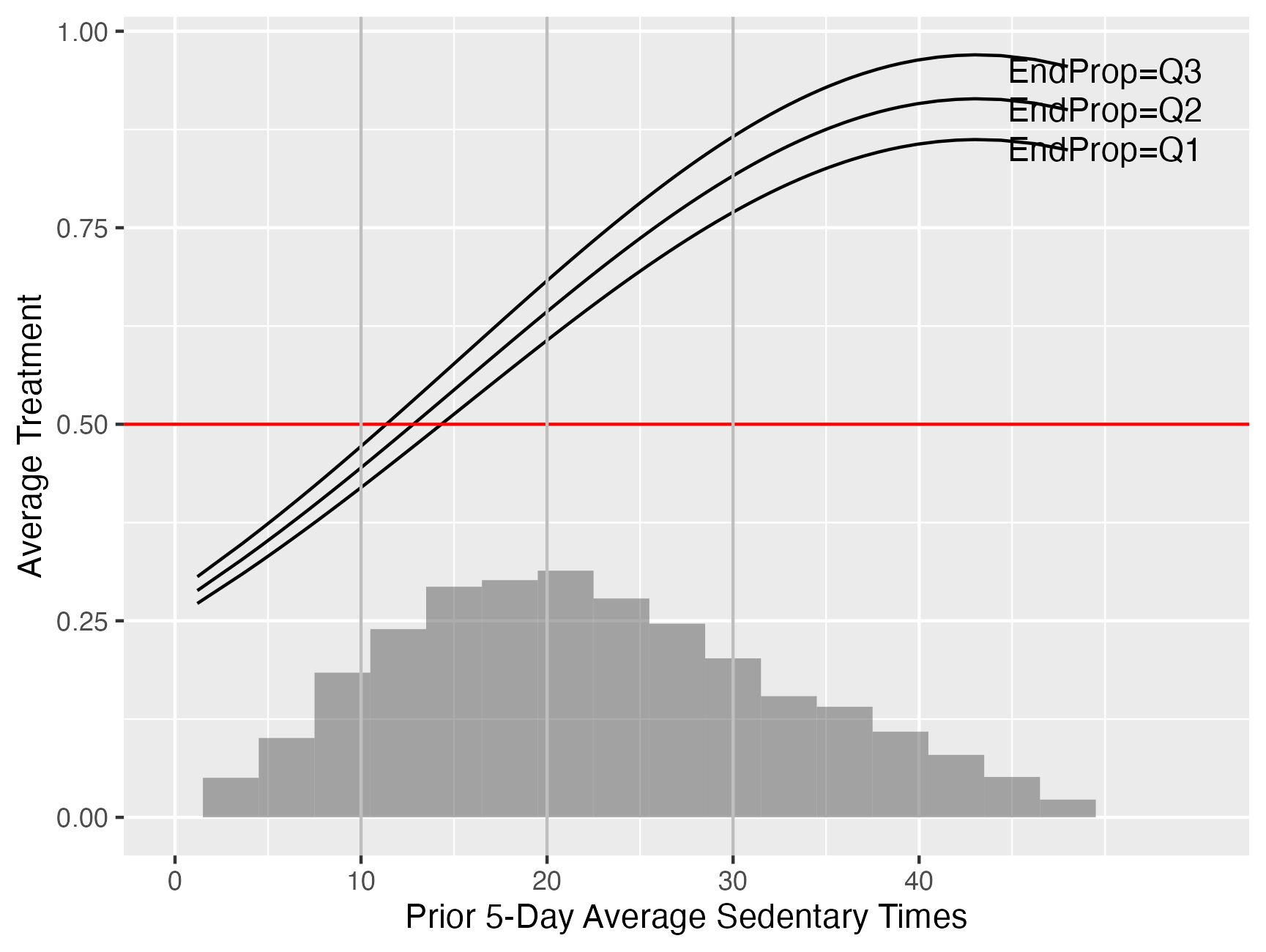}
    \includegraphics[width=0.4\linewidth]{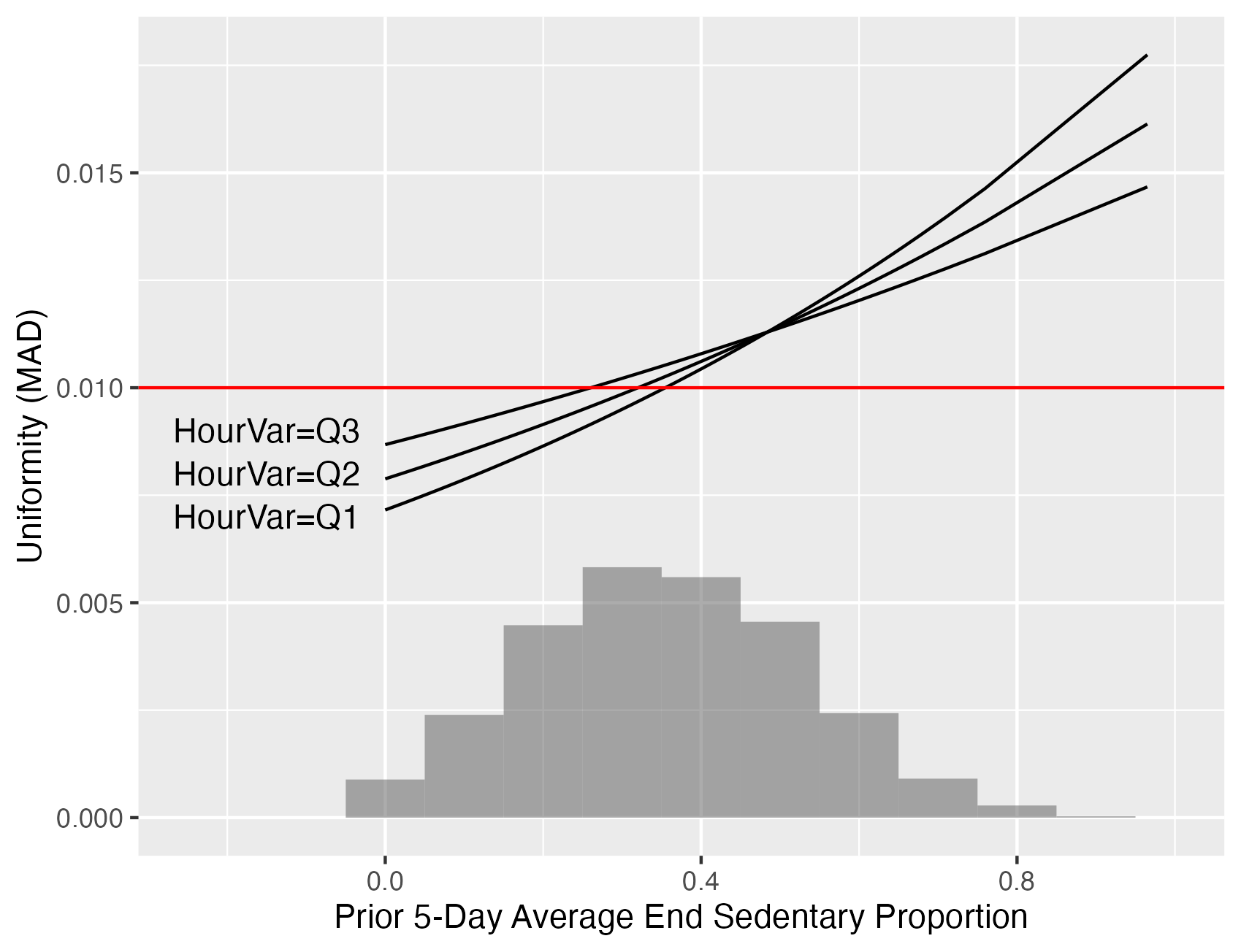}
    \caption{Predicted average treatment and MAD with respect to the most relevant features. Red horizontal lines are at 0.5 for average treatment and 0.01 for MAD, representing benchmarks to differentiate blocks with high and low average treatment and MAD values. Grey vertical lines represent plausible cutoffs to differentiate blocks. The three distinct lines in each plot represent the first quartile (Q1), median (Q2), and third quartile (Q3) of EndProp and HourVar values, respectively.}
    \label{fig:gee_fitted}
\end{figure}

\begin{table}[h]
\centering
\begin{tabular}{lccc}
\hline
Predictor & Estimate & Std. Err & p-value \\ \hline
Intercept & -0.543 & 0.037 & $<$ 2e-16 \\
MeanSed & 0.288 & 0.017 & $<$ 2e-16 \\
MeanSed\textsuperscript{2} & -0.083 & 0.013 & 1.95e-10 \\
EndProp & 0.529 & 0.095 & 2.56e-08 \\ \hline
\end{tabular}
\caption{Fitted parameters for time-varying covariates in the GEE regression on average treatment}
\label{table:gee-params}
\end{table}

The fitted parameters are displayed in Table \ref{table:gee-params}. In this table, the coefficient for MeanSed is presented after standardization. We refrain from interpreting these coefficient values here, as they specifically represent the performance of the current version of SeqRTS. These values are not generalizable and fall outside the primary focus of our discussion. Our main intention is to demonstrate the directionality of the relationships: there is a positive correlation between average treatment and both MeanSed and EndProp, which aligns with our reasoning at the beginning of the section.

We now identify blocks with different performances and provide a recommendation for improving the average treatment to achieve 0.5 treatments per block for each participant. To identify blocks with different performances, we plot the relationship between average treatment and MeanSed. The plot is shown on the left in Figure \ref{fig:gee_fitted}. We controlled for EndProp by setting it at the 25th, 50th, and 75th percentiles (Q1, Q2, Q3) to make three fitted lines. It is clear that EndProp makes a difference, but is not too influential as the three fitted lines are close. We focus on the line setting EndProp to its 50th percentile. 

A horizontal line representing the block average goal of 0.5 treatments is plotted. The points at which this line intersects with the fitted lines mark blocks where the algorithm’s performance deviates from its goal. A plausible categorization of the blocks in terms of different average treatments is as follows: \(\text{MeanSed} < 10\), \(10 \leq \text{MeanSed} < 20\), \(20 \leq \text{MeanSed} < 30\), and \(\text{MeanSed} \geq 30\). One strategy to improve the algorithm’s performance is to stratify blocks into these four categories. Before the next study batch, we could retrain the hyperparameters of the SeqRTS algorithm using the TUNE method described in \cite{liao2018just} for these four types of blocks. During the study, before the start of each day, we can use prior $K$-day (e.g. prior 5-day) average  of MeanSed as an input to determine the category of the block and then output the corresponding hyperparameters.

\subsubsection{Identify Time-varying Covariates to Improve Uniformity}
\label{sec:uniformity-gee}
In this section, we seek ways to improve uniformity by first identifying time-varying covariates that are correlated with uniformity, and then provide a recommendation based on the regression result. 

Note that MAD, the metric for uniformity, measures the differences among treatment probabilities within a block. To select time-varying covariates that may correlate with MAD, we focus on the factors influencing difference between consecutive probabilities. 
Let $t_k$ and $t_{k+1}$ denote two consecutive risk times, where \(t_{k+1} \geq t_k+1\) \footnote{Here, \(t_{k+1} \geq t_k+1\) but not strictly equal to \(t_k+1\) to account for the fact that the next risk time after $t_k$ might not be exactly the next decision time $t_k + 1$.}. We consider the strong influencers on the ratio between consecutive probabilities, \(\frac{p_{t_{k+1}}}{p_{t_k}}\), because when the ratio deviates from 1, variability in the whole set of probabilities in the block increases.
From the SeqRTS formula (Equation \ref{eq:p^{SeqRTS}}), we observe that
$$
p_{t_{k+1} } = \frac{N_{t_k} - p_{t_k}}{1 + \hat g_{t_{k+1}}} = \frac{N_{t_k} - N_{t_k} / (1+\hat g_{t_k})}{1 + \hat g_{t_{k+1}}} = \frac{N_{t_k}}{(1+\hat g_{t_k})} \frac{\hat g_{t_k}}{1 + \hat g_{t_{k+1}}} = p_{t_k} \frac{\hat g_{t_k}}{1 + \hat g_{t_{k+1}}}
$$
where $N_t = \hat N_0 - \sum_{s = 1}^{t-1} X_s \cdot p_s$ is remaining budget at time $t$ (recall in Section 3 that a budget is an approximate average treatment allowed in the block).  Moving $p_{t_k}$ from right to the left, we have
\begin{align}
\frac{p_{t_{k+1}}}{p_{t_k}} = \frac{\hat g_{t_k}}{1 + \hat g_{t_{k+1}}}. \label{eq:SeqRTS-prob-rate}
\end{align}

Equation \ref{eq:SeqRTS-prob-rate} indicates that the ratio between two consecutive treatment probabilities $\frac{p_{t_{k+1}}}{p_{t_k}}$ is characterized by the ratio between the estimated remaining sedentary times $\hat g_t$. Values of \(\hat{g}_t\) shown in Figure \ref{fig:hat_g_t} suggests two scenarios in which this ratio is large. First, when \(t_k\) and \(t_{k+1}\) are in the same hour, \(\hat{g}_{t_k}\) is almost equal to \(\hat{g}_{t_{k+1}}\). When they are equal, say \(\hat{g}_{t_{k+1}} = \hat{g}_{t_k} = s\), the ratio in Equation \ref{eq:SeqRTS-prob-rate} reduces to \(\frac{s}{1 + s}\). In this case, a smaller \(s\) leads to a larger ratio. Smaller \(s\) appears in the later hours of the block since \(\hat{g}_{t}\) is smaller in the later hours of the block (as shown in Figure \ref{fig:hat_g_t}). Note that EndProp defined in Section 5.2.1 describes the proportion of risk times in the later hours. We therefore use EndProp as a regressor and hypothesize that it positively correlates with MAD. Second, when \(t_k\) and \(t_{k+1}\) are in different hours, Figure \ref{fig:hat_g_t} shows that \(\hat{g}_{t_k}\) differs largely from \(\hat{g}_{t_{k+1}}\). As a result, both the ratio in Equation \ref{eq:SeqRTS-prob-rate} and MAD should be high when risk times appear in different hours. To capture this phenomenon, we define HourVar (hour variation) as the number of distinct hours during which risk times occur in a block and use it as a regressor.

We now conduct a GEE regression analysis with EndProp and HourVar as two covariates and MAD as the outcome. 
An exploratory analysis indicates that $\text{EndProp}$, $\text{HourVar}$, $\text{HourVar}^2$, and $\text{EndProp} \times \text{HourVar}$ are correlated with the outcome. When fitting a GEE model with an AR-1 working correlation model, we found that $\text{HourVar}^2$ is not significant. We refine the GEE model to exclude $\text{HourVar}^2$. This refined model has Pearson residuals around zero, suggesting its plausibility. The coefficients for EndProp, HourVar, and their product are 2.20, 0.27, and -0.52, respectively, and all are significant.

With fitted regression model, we now identify blocks for which MAD is large. We plot the average MAD against EndProp, holding HourVar at its three quantiles. We use MAD = 0.01 as the benchmark to differentiate blocks with high and low MAD, as argued in Section 4. The plot is shown on the right of Figure \ref{fig:gee_fitted}. We observe the following from the plot: First, HourVar is not impactful since the three fitted lines look similar. Second, under the same HourVar, a large EndProp positively correlates with a large MAD.

These findings suggest that one way to improve uniformity is to gradually decrease \(\hat{g}_t\) over decision times in a block. This approach aims to mitigate the impact of EndProp: as discussed above, the reason EndProp leads to high MAD is that \(\hat{g}_t\) remains almost the same when two risk times are within the same hour. By decreasing \(\hat{g}_t\), i.e., ensuring \(\hat{g}_{t_{k+1}} < \hat{g}_{t_k}\) for two consecutive risk times \(t_k\) and \(t_{k+1}\), instead of keeping them the same, the ratio \(\frac{p_{t_k}}{p_{t_{k+1}}} = \frac{\hat{g}_{t_k}}{1 + \hat{g}_{t_{k+1}}}\) will be closer to one, thereby reducing the MAD in the block. One simple strategy to achieve this is to estimate the proportion of sedentary times in a block (sedentary rate) and then let \(\hat{g}_t\) be the product of the sedentary rate and the remaining decision times in the block. Other strategies might better improve uniformity, which we leave for future work.

\section{Conclusions and Discussion}
\label{sec:conclusion}

A new and challenging online decision problem has emerged in the evolving field of MRTs: determining a treatment probability that adheres to an average treatment constraint and a uniformity constraint when the times at which randomizations are allowed (risk times) are random. HeartSteps V2V3 MRT (HSV2V3) implemented the SeqRTS algorithm to meet these design constraints. Analyzing data from the HSV2V3, we found that SeqRTS performed promisingly in adhering to the core design constraints in the real world: the average treatment approximated the target of 1.5 per day, and the uniformity constraint was satisfied well on average across participants. We now discuss key considerations in evaluating, improving, and implementing the SeqRTS algorithm, in future studies.

\subsection{Handling Missing Data in the Evaluation}
In the evaluation, data for the algorithm's performance was missing at unavailable decision times and on unavailable days because the algorithm could not perform during those times and days. Below, we discuss how the missing data are handled in our analysis. 

First, it is important to clarify both the assumptions and the target for the evaluation when excluding days on which a participant is unavailable. We have concluded that days we excluded were either days on which we are not interested in assessing the algorithm’s performance (days when a user is active or does not wear the Fitbit) or days on which we are interested (days when a user wears the Fitbit but there is no connection to the server), under which the data's missingness is at random (MAR). As a result, the target for the evaluation is the algorithm's performance on days when participants wear the watch and are not active all day, and the analysis using only the complete data does not pose a bias issue.

One potential concern of the result above is the influence of unobserved variables, such as mental tiredness (\(U\)), which may affect both a participant's decision to wear the watch and the accuracy of the algorithm's predictors, like \(\hat{g}_t\). As a result, the algorithm may seem to perform well by excluding days when participants do not wear the watch, potentially masking underlying challenges. We recognize that our analysis is based on a population representative of Kaiser Permanente patients newly diagnosed with stage 1 hypertension, and the algorithm's performance is evaluated marginally across variables like \(U\).

However, as patient populations evolve, the marginal effects may also change. Future studies should consider periodically updating the SeqRTS algorithm to ensure it remains effective in light of shifting behavior patterns.

Second, when handling unavailable decision times in a day, we assume that a participant does not respond to the anti-sedentary message, i.e., the anti-sedentary message does not affect a participant's availability. This assumption is important for ensuring the correct interpretation of the evaluation results and the recommendations for the algorithm's improvement, as otherwise, a participant's responsiveness will entangle them. We illustrate this issue using the following example. Consider two users who differ in both sedentariness and responsiveness to the algorithm: User A is very sedentary and very responsive to messages. Being very sedentary means there is a high chance that one message is sent in a day, say  80\%. Being very responsive means once they receive one message, they become active and not available for the rest of the day. This results in an average of 0.8 messages per day. User B is less sedentary but not responsive to the messages at all. Because User B is less sedentary, there is a lower chance of receiving one message in a day, say 50\%. But because they do not respond to the message, there is also a chance, say 15\%, that they receive two messages in a day. This results in an average of \(1 \times 50\% + 2 \times 15\% = 0.8\) messages per day (assuming User B receives zero message on the remaining 35\% of days). This example indicates that the algorithm’s performance on a given day could be impacted by both a user's sedentariness and responsiveness. We assume users do not respond to the anti-sedentary message, making sedentariness the sole influencer of the algorithm's performance. It will be interesting for future works to explore methods to disentangle sedentariness and responsiveness.

\subsection{The Metric for Evaluating Uniformity}
Our key methodological contribution to assessing the algorithm's performance is the proposal of mean absolute deviation (MAD) as a fairer metric for measuring uniformity than KL divergence, as proposed in \cite{liao2018just}. This is because, in addition to measuring the algorithm's performance on meeting the uniformity constraint, KL divergence also depends on its performance on the average treatment constraint. Poor performance on the average treatment constraint can result in a high KL value, even when the uniformity constraint is well met. In contrast, MAD measures SeqRTS's performance on the uniformity constraint alone and is not influenced by the algorithm's performance on the average treatment constraint.

\subsection{Ways to Improve the Algorithm’s Performance}
With the ultimate goal of meeting two constraints for each participant, we have identified baseline and time-varying covariates that correlate with the algorithm's performance and quantified their relationship to the algorithm's performance. Based on the findings, we offer the following suggestions to improve the SeqRTS algorithm.

First, to improve uniformity, we should gradually decrease \(\hat{g}_t\) over decision times in a block, instead of keeping \(\hat{g}_t\) values nearly constant within the same decision hour, as done in HSV2V3. In Section \ref{sec:uniformity-gee}, we showed that this ``constant" pattern causes large variations in treatment probabilities for consecutive risk times within the same hour. One simple strategy to gradually decrease \(\hat{g}_t\) is to estimate the proportion of sedentary times in a block (sedentary rate) and multiply it by the number of remaining decision times in the block.\footnote{This approach gradually decreases \(\hat{g}_t\) because the sedentary rate is constant, but the remaining decision times in the block are deterministic and decrease with \(t\).} There could be other strategies that further improve uniformity. We leave this as a future work.

Second, to improve average treatment, we should use the degree of sedentariness to categorize participants and blocks and retrain the algorithm's parameters for these categories. At the participant level, in Section \ref{sec:avg-trt-bsl}, we found that participants who are highly sedentary during the baseline period are over-treated, those who are highly active during the baseline period are under-treated, and participants in the middle ground are well-treated. We suggest improving treatment averages by training three sets of hyperparameters for SeqRTS using data from these three types of participants. One way to train the hyperparameters is the \texttt{TUNE} method described in \cite{liao2018just}. Once the hyperparameters are trained, one can use baseline variables in the next study to classify participants into the above three categories and assign them the corresponding hyperparameters to warm-start the study.

At the block level, in Section \ref{sec:avg-trt-gee}, we found that the prior 5-day average sedentariness (MeanSed) is positively correlated with the number of treatments. Specifically, blocks with \(\text{MeanSed} < 10\) are under-treated. Blocks with \(10 \leq \text{MeanSed} < 20\) are well-treated. Block with \(20 \leq \text{MeanSed} < 30\) are over-treated, and blocks with \(\text{MeanSed} \geq 30\) are too-overly-treated. As above, we could train the hyperparameters of the SeqRTS algorithm using TUNE for these four types of blocks. We recommend use MeanSed  before the start of each block during the study as an input to determine the block's category and then apply the appropriate hyperparameters. During the study, before the start of each block, the prior \(K\)-day (e.g., prior 5-day) average of MeanSed can be used as an input.

\subsection{Considerations for Real-World Implementation and Data Quality in SeqRTS}
We conclude this case study with several considerations for the future implementation of the algorithm in the real world, aiming for easier and more accurate evaluation of the algorithm. First, code testing for the algorithm is crucial, as the cost of coding issues can be substantial. During the evaluation phase, we encountered a coding issue that significantly impacted the evaluation process, as detailed in Appendix \ref{sec:coding-issue}. The issue resulted in the algorithm being unable to perform as intended (see Section 4.1 for a discussion). Although this issue affected a small proportion of the data, it had a considerable impact: including the affected data would improperly represent the algorithm’s performance on average treatment and uniformity. Consequently, we had to remove the affected data from the analysis, which was a loss.

Second, we need to enhance the quality of data collection during baseline period. In mobile health studies with MRTs, it is essential to consider post-study analysis when researchers collect data \citep{zhang2022statistical}. In Section 5.1, we highlighted that the significance of sedentariness and activity level variables collected from the baseline survey influence the algorithm's performance, and we recommend researchers use them to improve the algorithm in future studies. However, the sedentariness data for 9 of the 82 participants was missing. Although we conducted sensitivity analyses using several imputation methods, it is always better when the data is not missing. Therefore, it is imperative to improve the quality of data collection concerning sedentariness and activity level at baseline in future studies.

Third, it is crucial to keep data used by the decision algorithm independent during a study. Specifically, in HSV2V3, we defined unavailability using four criteria but resorted to only one indicator when all availability criteria were met, instead of keeping four indicators, one for each availability criterion. The third unavailability criterion—unavailability within one hour after any type of notification is sent—created dependence for the sole availability indicator on the algorithm implemented in the study. As a result, future testing of different algorithms based on this study data will be unreliable. This issue is also pertinent when we find time-varying variables that correlate with the algorithm's performance, as they all depend on the risk indicator, and hence the availability indicator. In the next iteration of SeqRTS, we suggest keeping four indicators, one for each availability criterion. This suggestion should also apply to all stratified micro-randomized trials where risk times are of concern in decision making \citep{dempsey2020stratified}.

\section*{Acknowledgments}

This research was supported by several grants. We acknowledge funding from the following sources: NIH/NIDA P50 DA054039, NIH/NIBIB and OD P41EB028242, NIH/NHLBI R01HL125440, and NIH R01 GM152549.

\bibliography{references}

\newpage

\appendix 
\section*{Appendix}

\section{The Sequential Risk Time Sampling Algorithm Parameters Used in the HS V2V3 Studies}
\label{appdix:hat-g}

In this section, we explain how we set tuning parameters for the SeqRTS algorithm. First, we explain the rationale behind setting the upper bound of treatment probabilities to 0.2. Then, we describe the model for \(\hat{g}_t\). After both were chosen, \(\hat{N}_0\) was tuned using the TUNE algorithm from \cite{liao2018just}. Finally, we present how well \(\hat{g}_t\) performed using data from HSV2V3.

\subsection{Determine the Upper Bound of the Treatment Probabilities}
\begin{figure}
    \centering
    \includegraphics{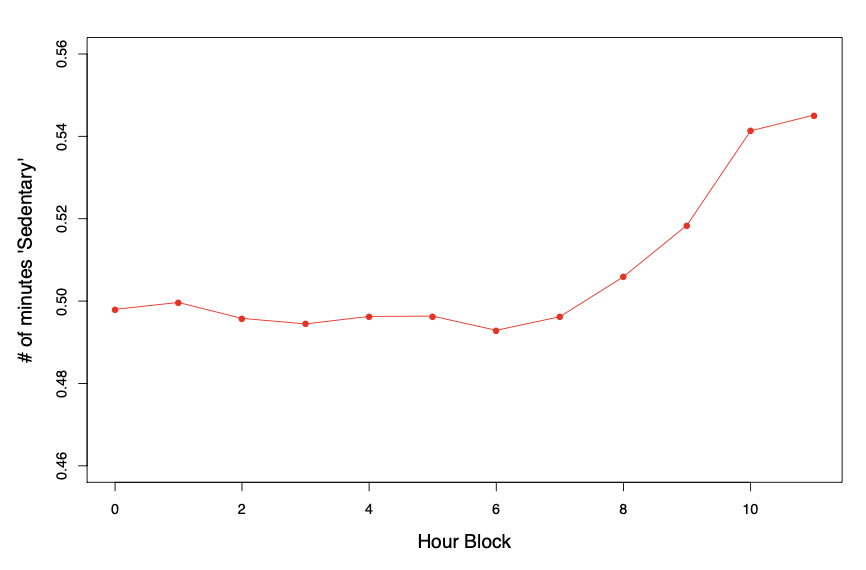}
    \caption{Fraction of time sedentary within and after current hour}
    \label{fig:frac-sed-V1}
\end{figure}

In this section, we explain the rationale behind setting the upper bound of treatment probabilities to 0.2. Figure \ref{fig:frac-sed-V1} illustrates the fraction of time individuals remain sedentary during different hours of the day. For instance, at 10 AM, the fraction of sedentary time is calculated from 10 AM to 9 PM. This calculation is repeated hourly from 9 AM to 9 PM (i.e., at 9 AM, 10 AM, ..., 8 PM, 9 PM). The purpose of this computation is to predict sedentary behavior for the remainder of the day, rather than just for the upcoming hour. The plot shows a relatively consistent range, fluctuating between 0.50 at the start of the day and 0.55 towards the end. Consequently, a randomization probability of 0.2 (excluding availability constraints) results in an average of approximately 144 × 0.525 × 0.2 = 15.12 treatments per day. This upper limit is chosen to prevent the risk of under-treatment. The intention is to circumvent scenarios where only a few sedentary periods occur towards the end of the day. In such cases, it is crucial to avoid excessively high probability of intervention. Setting the probability at 0.2 ensures that, even with as few as 10 sedentary periods in a day, the average treatment constraint is likely to be met.

\subsection{Model for $\hat g_t$}

\begin{figure}
    \centering
    \includegraphics[width=0.9\linewidth]{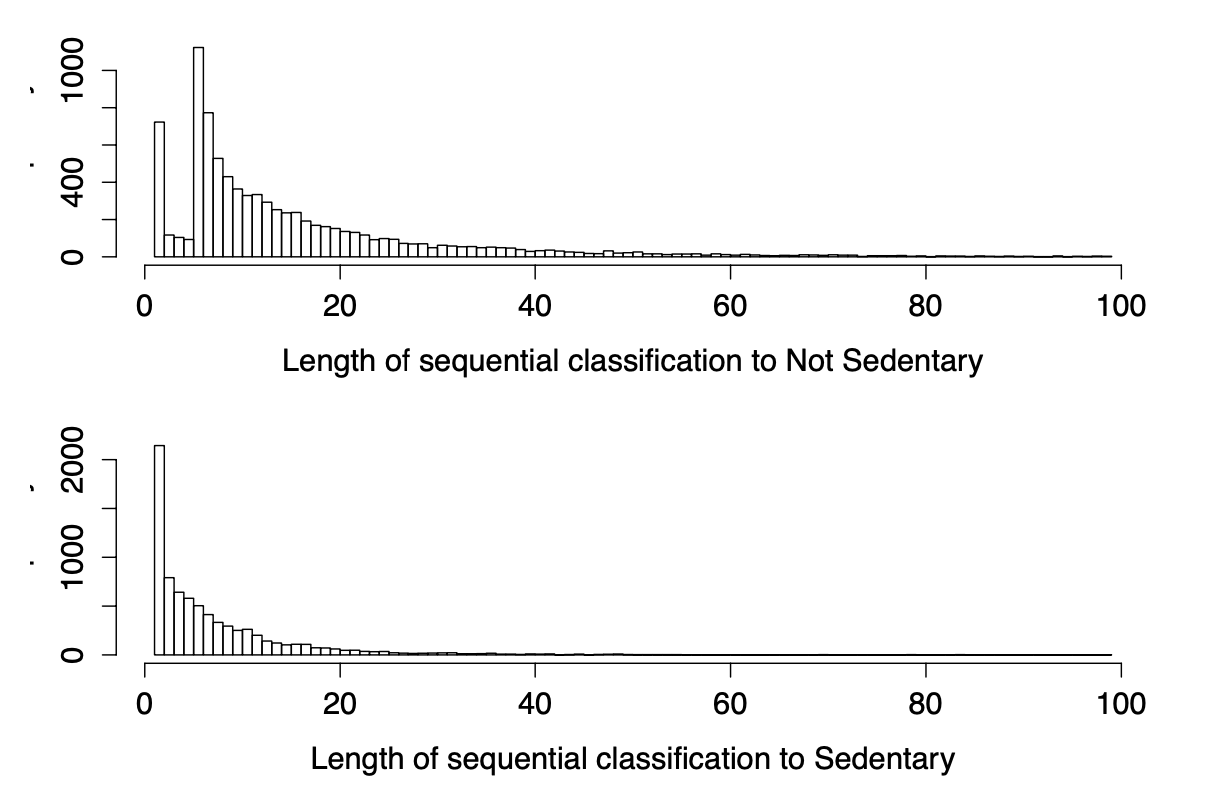}
    \caption{Run lengths of being classified as either “Not Sedentary” or “Sedentary” respectively.}
    \label{fig:hist-of-run-length-V1}
\end{figure}

Now we explain our choice of model for $\hat g_t$. Given the current run length $R_t$, define $K$ to be the remaining amount of sedentary ``run to go" in the current state (so $R=K+R_t$ is the total run length). We then define $F$ to be the fraction of time in the state "Sedentary" remaining in the day, as shown in Figure \ref{fig:frac-sed-V1}. Then the amount of remaining time the participant is classified as "Sedentary" is given by $\Omega=K \wedge r+F(r-K)_{+}$, where $\wedge$ denotes minimum operator and $(x)_{+}=\max (x, 0)$ and $r$ denotes the amount of time remaining in the current day. That is, we account for the fact that the expected run length may be greater than the remainder of the day. We assume that $F$ is independent of both functionals of $K$ given $k$ and $B_s$. The conditional expectation can then be written as
$$\mathbb{E}\left[\Omega \mid R_s, B_s\right]=\mathbb{E}\left[K \wedge r \mid R_s, B_s\right]+\mathbb{E}\left[F \mid R_s, B_s\right] \cdot \mathbb{E}\left[(r-K)_{+} \mid R_s, B_s\right]$$
Below we assume $B_s=1$ as we only intervene when the individual is classified as "Sedentary". Each component of the conditional expectation is estimated using the Heartsteps V1 data. Figure \ref{fig:hist-of-run-length-V1} is a Histogram of run lengths in state "Sedentary" and "Not Sedentary" respectively. We checked the Heartsteps V1 data for differences as functions of “time of day” and “length in study”, but found very little to suggest a more complicated model was necessary. 

We first estimate $\mathbb{E}\left[K \wedge r \mid R_s = k, B_s = 1\right]$ and $\mathbb{E}\left[(r-K)_{+} \mid R_s = k, B_s = 1\right]$. Let $\left\{\tilde{R}_{i}\right\}_{i=1}^{m}$ denote the data of run lengths used to compute the ``Sedentary" plot within Figure \ref{fig:hist-of-run-length-V1} as $\tilde R_i$. For each $k = 1, 2,.., 96$ (two times length of a block), define $\tilde{K}_{i}=\tilde{R}_{i}-k .$ Using only runs such that $\tilde{K}_{i} \geq 0$, we have the following estimates
$$
\begin{gathered}
\mathbb{E}\left[K \wedge r \mid k, B_s\right] \approx \frac{1}{m_{k}} \sum_{j=1}^{m_{k}} \tilde{K}_{j} \wedge r \\
\mathbb{E}\left[(r-K)_{+} \mid k, B_s\right] \approx \frac{1}{m_{k}} \sum_{j=1}^{m_{k}}\left(r-\tilde{K}_{j}\right)_{+}
\end{gathered}
$$
where $m_{k} \leq m$ is the number of observations satisfying the condition $\tilde{R}_{i} \geq k$.

We then estimate the fraction of time classified as Sedentary in the remaining window, $\mathbb{E}\left[F \mid k, B_s\right]$. Figure \ref{fig:frac-sed-V1} plots the estimate calculated using the data from the current hour and all remaining hours. Here, since the figure shows this fraction is flat, we simply use the current hour estimate. For example, at time 9 : 15AM, we use the estimated fraction in Figure \ref{fig:frac-sed-V1} from 9 : 00AM—i.e., the fraction of time sedentary from 9 : 00AM to 10 : 00PM. At 9 : 45AM, we use the same estimate; however, at 10 : 00AM and 10 : 10AM, we use the estimated fraction at 10 : 00AM—i.e., the fraction of time sedentary from 10 : 00AM to 10 : 00PM.

\subsection{Quality of $\hat g_t$}
\label{sec:quality-hatgt}

This section evaluates the overall effectiveness of $\hat g_t$ in predicting the actual remaining risk times $g_t$. The primary finding is that $\hat g_t$ tends to significantly overestimate. We calculated the Root Mean Square Error (RMSE) for each block between $\hat g_t$ and $g_t$. The average RMSEs are presented in Table \ref{tab:sed_MAD}. According to the table, for a median user, $\hat g_t$ was overestimated by an average of 3 across all risk times in each block. Considering that a user typically has an average of 7 risk times, this overestimation is substantial.

\begin{table}[ht]
	\centering
	\caption{Mean and Mean Absolute difference (MAD) for HSV1 and HSV2V3 — from Liao et al (2018)}
	\label{tab:sed_MAD}
	\begin{tabular}{ccc}
		\toprule
		& \multicolumn{2}{c}{HSV2V3} \\
		\cmidrule(r){2-3} 
		& Mean Risk Times & RMSE of $\hat g_t$ \\
		\midrule
		Block 1 & 7.79 & 3.24 \\
		Block 2  & 7.90 & 3.23 \\
		Block 3  & 8.28 & 3.17 \\
		\bottomrule
	\end{tabular}
\end{table}

\section{A Coding Issue that Affects Evaluation}
\label{sec:coding-issue}
In analyzing HeartSteps V2V3, a coding issue in the algorithm was discovered. This issue makes the SeqRTS algorithm unable to perform as intended. As a result, the algorithm's performance on both average treatment and uniformity was impacted. We discuss the impact of the coding issue on average treatment and uniformity, the exclusion criteria for data that is largely impacted by this issue, and how it led us to choose data for analysis.

This issue is due to an incorrect conversion of the timezone. There is a need for timezone standardization because participants have different start times and may live in or travel to different timezones, though most participants lived in Seattle. The code attempts to standardize all start times to a reference time of 8am EST because 8am typically marks the beginning of most participants' day.

Let \( t^{GMT} \) be a decision time represented in GMT and \( t^{EST} \) be the same time represented in EST. Let \( t_0^{GMT} \) represent the starting time in GMT. GMT is involved because it is the default timezone used by R, the coding language that the algorithm employs. The conversion aims to obtain \( t^{EST} \) using the formula \( t^{EST} = t^{GMT} - (t_0^{GMT} - 14) \), where 14 is chosen because 8am EST corresponds to 2pm (or 14:00) in GMT. This sometimes leads to values outside the expected range when a participant's actual start time deviates from 8am EST, making the \(\hat{g}_t\) parameter undefined.

For instance, if a participant begins at 5pm GMT (12pm EST), then at 1am GMT, instead of it being mapped to the correct value 7pm EST, we end up with a calculation of \( 1am - (17pm - 14pm) = -2 \). This makes \(\hat{g}_t\) a NA value in the code. This issue occurs in a small portion of block 2 and a significant portion of block 3. All probabilities at these times are set to the algorithm's lower bound of 0.005.

We use the variable \texttt{invalid\_param\_proportion}, mathematically denoted as \(IPP\), as the proportion of risk times when the SeqRTS's parameter becomes invalid due to the coding issue, i.e., the proportion equals to the number of ``impacted risk times" divided by 48. Let $k > 0$ be a given threshold value of the proportion.  Blocks can be categorized into  either $IPP < k$ or $IPP \geq k$. The proportion of the latter category blocks among all blocks is denoted by \texttt{Proportion\_data\_GE\_k} (proportion of data greater or equal to k). Analyzing the impact involves constructing confidence intervals for the average treatment and uniformity between the two data categories. We vary $k$ from 0.1 to 0.9. For each fixed $k$, we present point estimates and confidence intervals of the differences in average treatment and and uniformity separately in  Table \ref{tab:results}. We also show the proportion of blocks whose $IPP < k$.

\begin{table}[h]
\centering
\begin{tabular}{|c|c|c|c|}
\hline
$k$ & Proportion\_data\_GE\_k & Difference in No. Treatments & Difference in Uniformity\\
\hline
0.1 & 0.157 & -0.109 [-0.136, -0.081] & -0.305 [-0.313, -0.297] \\
\hline
0.3 & 0.1 & 0.046 [0.015, 0.077] & -0.336 [-0.347, -0.325] \\
\hline
0.5 & 0.054 & 0.310 [0.279, 0.341] & -0.306 [-0.325, -0.286] \\
\hline
0.7 & 0.03 & 0.434 [0.403, 0.465] & -0.157 [-0.187, -0.126] \\
\hline
0.9 & 0.018 & 0.505 [0.480, 0.531] & 0.112 [0.089, 0.135] \\
\hline
\end{tabular}
\caption{Comparisons between blocks where \(IPP \geq k\) and \(IPP < k\). "Proportion\_data\_GE\_k" represents the data proportion with \(IPP \geq k\). The point estimate and 95\% CI for the difference for these data subsets are presented. }
\label{tab:results}
\end{table}

The difference between blocks with low \(IPP\) and those with high \(IPP\) shows that the coding issue indeed influences the algorithm's performance. We choose to exclude entries where \(IPP \geq 0.5\) since they introduce a substantial discrepancy between average treatment and uniformity and represent only a small fraction of the dataset.

\section{Checking Plausibility of the MCAR Assumption}
\label{sec:MAR-vs-MCAR}

Note that the evaluation is only conducted on days and blocks when data is not missing for the entire day or block. We acknowledge that the validity of the GEE results relies on the assumption of missing completely at random (MCAR). In this section, we show that this assumption is plausible.

MCAR means the missingness does not depend on any observed variable. However, it might be plausible that some variables can predict the missingness. Two such variables are the weekend and the previous day’s number of available times. The weekend might be predictive since it is possible that on weekends, a user has low availability due to not wanting to wear the watch or going on a hike. The previous day’s availability might predict the current day's availability because the more that the user is available on the previous day, the more likely it is that the user remembers or wants to wear the watch the next day, i.e., the likelihood they are available is greater.

We use a regression model to check the whether these two varibles can predict the missingness. We choose GEE with a logistic link and AR-1 working correlations. The mean model is as follows: 
\begin{equation}
	\text{logit}P[I_d = 1] = \beta_0 + \beta_1 N_{d-1} + \beta_2 \text{is\_weekend}_d
\end{equation}
The results show that \(\beta_1\) is estimated to be 0.0164 with a p-value less than \(10^{-5}\), and \(\beta_2\) is not significant at 0.05 level. This indicates that the more the user is available on the previous day, the more likely it is that the user is available the next day. The weekend variable does not affect availability. 

The above says missing at random (MAR) might be more plausible since there is one observed variable that can predict the availability. To account for MAR, we use weighted GEE (WGEE, \cite{robins1995analysis}) to estimate the marginal performance of the algorithm. The result is presented in Table \ref{tab:gee-wgee-marginal}. 

The weights used in WGEE are left skewed, with the 1st Quartile at 1.1559, the Median at 1.2002, and the 3rd Quartile at 1.2357. These weights indicate minimal variation in the probability of missingness, suggesting that the observed data is nearly uniformly representative of the full dataset. Consequently, the WGEE results are very close to those obtained from the non-weighted GEE. This indicates that the estimates are robust to the assumption of MAR and do not significantly differ from those under the MCAR assumption. Therefore, we conclude that the impact of the potential MAR mechanism is minimal in our context.

Based on this finding, we proceed with the MCAR assumption in the main paper for simplicity and clarity. This approach allows us to use the non-weighted GEE results without a substantial loss of accuracy or validity.

We have identified two kinds of missing days we are not interested in, but there is one kind of missingness we are interested in (when a participant wears the watch, but there is no connection to the server). In this analysis, we do not differentiate the types of missing days (mainly because we are unable to do so), but we compare MCAR and MAR across all kinds of missing days. If we were able to differentiate the types of missing days, we would analyze a subset of the data.

\begin{table}[ht]
	\centering
	\caption{Average Treatment and Uniformity using GEE and WGEE }
	\label{tab:gee-wgee-marginal}
	\begin{adjustbox}{width=\textwidth}
		\begin{tabular}{llcccc}
			\toprule
			& &  \multicolumn{2}{c}{Avg. Trt.} & \multicolumn{2}{c}{Uniformity} \\
			\cmidrule(lr){3-4} \cmidrule(lr){5-6}
			Block / Day & Measure & AR-1 & AR-1 (WGEE) & AR-1 & AR-1 (WGEE) \\
			\midrule
			Whole Day & Mean & 1.534 (1.408, 1.659) & 1.521 (1.396, 1.647) & 0.019 (0.018, 0.020) & 0.019 (0.018, 0.020) \\
			& AR-1 Correlation & 0.306 (0.032) & 0.306 (0.032) & 0.163 (0.021) & 0.167 (0.022) \\
			Block 1 & Mean & 0.553 (0.479, 0.625) & 0.551 (0.478, 0.625) & 0.010 (0.009, 0.011) & 0.010 (0.009, 0.011) \\
			& AR-1 Correlation & 0.135 (0.019) & 0.135 (0.019) & 0.055 (0.016) & 0.056 (0.016) \\
			Block 2 & Mean & 0.540 (0.462, 0.618) & 0.539 (0.462, 0.617) & 0.011 (0.011, 0.012) & 0.011 (0.011, 0.012) \\
			& AR-1 Correlation & 0.139 (0.026) & 0.139 (0.026) & 0.016 (0.018) & 0.016 (0.018) \\
			Block 3 & Mean & 0.547 (0.475, 0.618) & 0.548 (0.477, 0.620) & 0.016 (0.015, 0.018) & 0.016 (0.015, 0.018) \\
			& AR-1 Correlation & 0.142 (0.021) & 0.142 (0.021) & 0.176 (0.027) & 0.177 (0.027) \\
			\bottomrule
		\end{tabular}
	\end{adjustbox}
\end{table}\footnote{Note that the usual non-weighted GEE coefficient estimates are different from the ones in the main paper. This is because the dataset we use here is a subset of the full dataset to ensure the previous day of the analysis day is available (so the weight, one over the probability of being available, can be estimated). }

\section{Checking the assumption that the mean is constant across the days}
\label{sec:no-day-trend}
In this section, we check the assumption that the mean is constant across the days. We justify this by showing no day trend for both the average number of treatments and the average MAD using both exploratory data analyses (EDA) and tests of significance.

We conduct an exploratory analysis by plotting $\bar{Y}_{j,d}=(1/n)\sum_{i=1; d\in \mathcal{D}_i}^{82} Y_{i,j,d}$, $j = 0,1,2,3$. This is the average of number of treatments on day $d$ in block $j$, averaged over users who are available on the day. Given the assumption that data is MCAR (see Appendix \ref{sec:MAR-vs-MCAR}), $\bar{Y}_{j,d}$ is a good proxy of the true population average number of treatments on day $d$. Thus, inspecting $\bar{Y}_{j,d}$ over days $d$ is a way to inspect whether the population mean number of treatment has day trend. On each plot we fix block number $j$ and plot $\bar{Y}_{j,d}$ over days $d$. We do the same for all blocks separately. 

\begin{figure}
    \centering
    \includegraphics[width=0.5\linewidth]{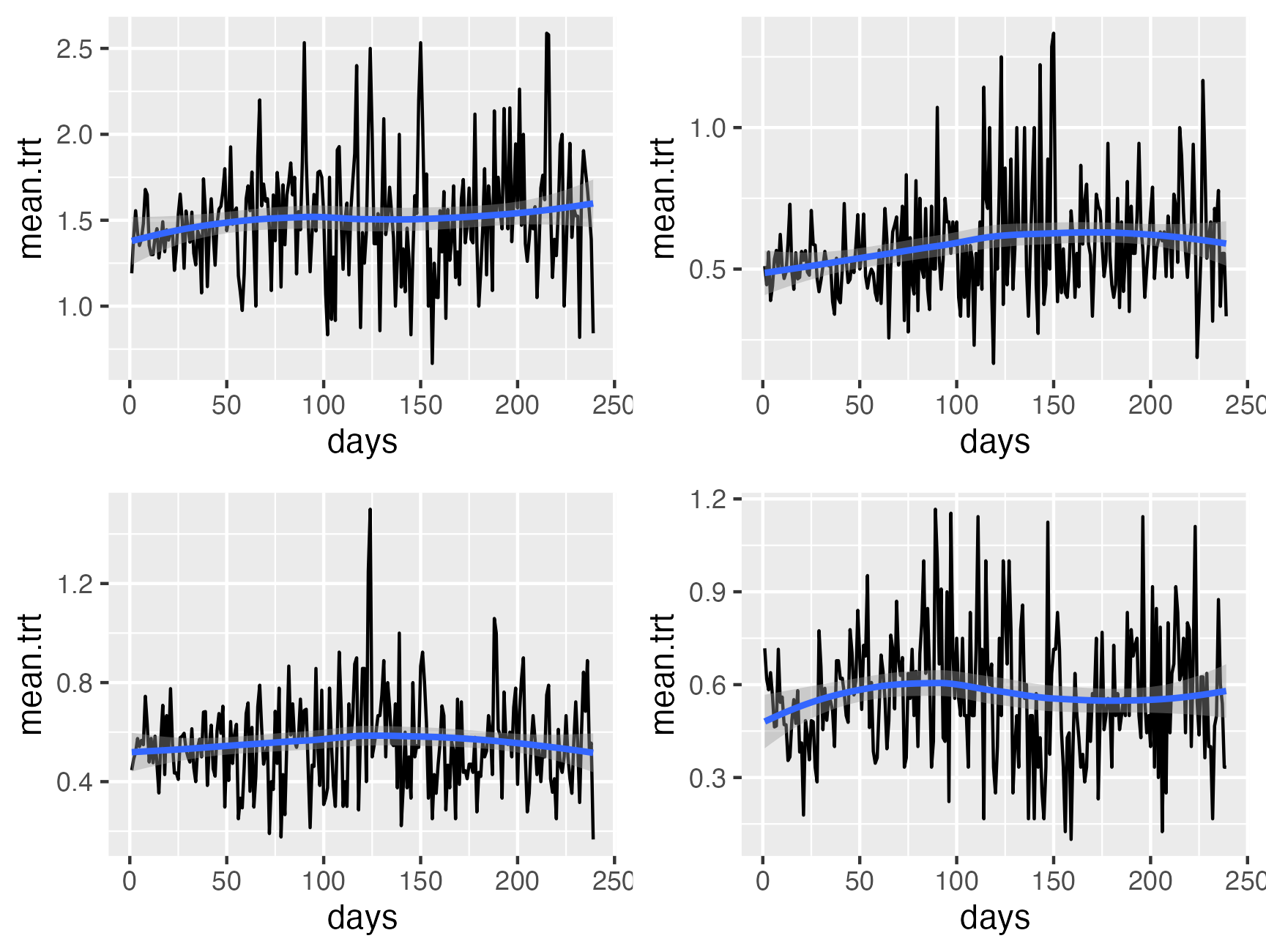}
    \caption{Plot of average of number of treatments over all study days for all blocks. The averages are taken over users who are available on the day.}
    \label{fig:EDA-Day-n-trt-Plot}
\end{figure}

Figure \ref{fig:EDA-Day-n-trt-Plot} shows the plot. The variation is larger between day 100 to day 150 because the number of available users in those days are low (due to the server issue described in Section 2). The plot shows that in general, there is no variation or mildly increasing trend across days.

To validate the hypothesis, we conducted the regression analysis below. This model is a GEE model specified in Equation \ref{eq:GEE-avg-trt} with the study day term ($\text{Day}_d$) added. 
\begin{align}
   &  \log E[Y_{i,j,d}|X_{i,j,d}] =\log \mu_{j} + \gamma \text{Day}_d \label{eq:GEE-avg-trt-by-bsl}  \\ 
   & Var(Y_{i,j, d}|X_{i,j,d}) =  \eta E[Y_{i,j,d}|X_{i,j,d}] 
\end{align}
Here, $\gamma$ represents the day trend. We perform the analysis for each block $j=0,1,2,3$. All results show that $\gamma$ is not significant at 0.05 level, indicating there is no day trend. 

\begin{figure}
    \centering
    \includegraphics[width=0.5\linewidth]{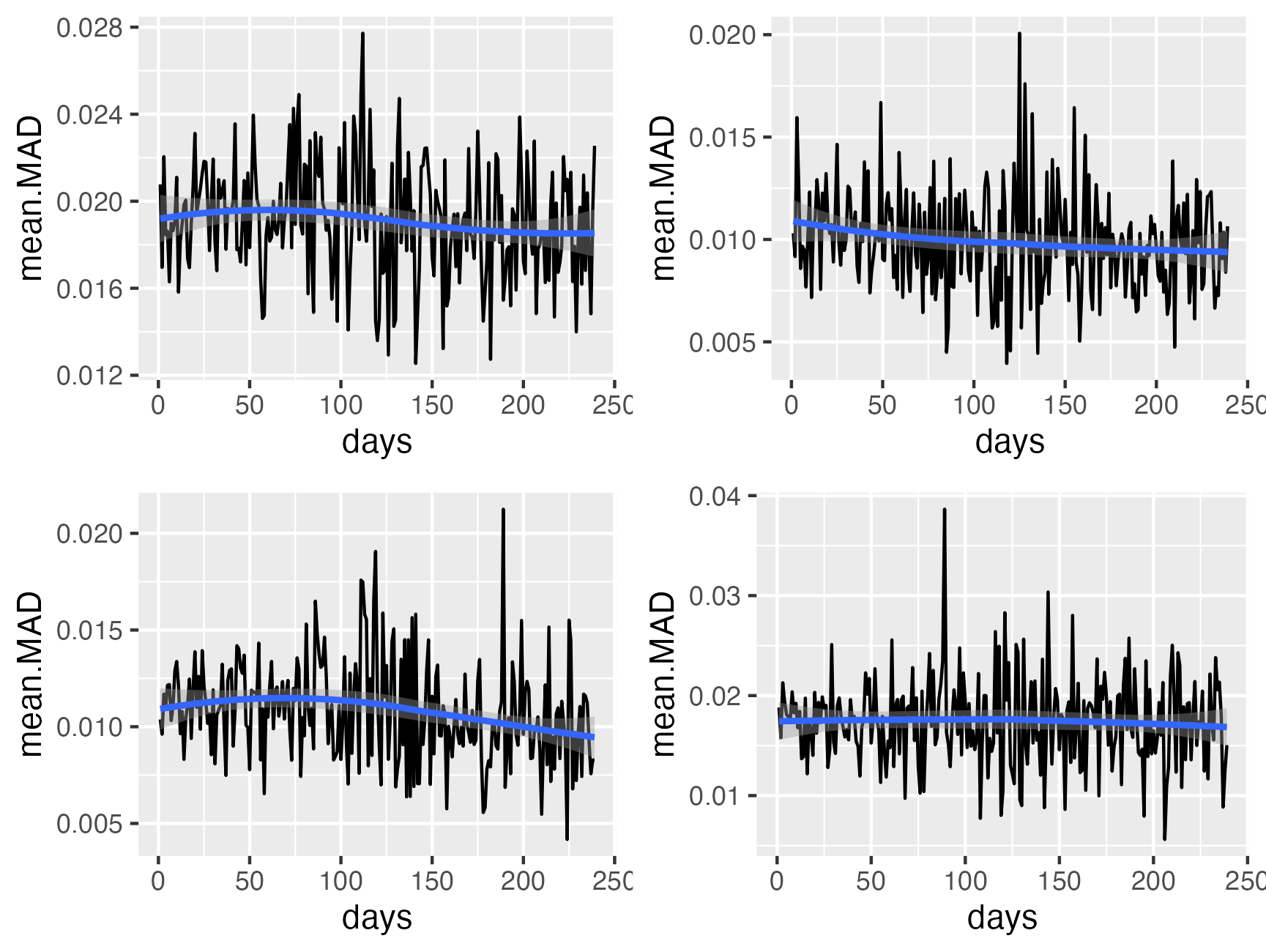}
    \caption{Plot of average MAD over all study days for all blocks. The averages are taken over users who are available on the day.}
    \label{fig:EDA-Day-MAD-Plot}
\end{figure}

For MAD, the plots (shown in Figure \ref{fig:EDA-Day-MAD-Plot}) are generally stable, too. Block 1 shows a slightly decreasing trend. However, when we fit a study day term into the regression, the regression coefficient is not significant, indicating there is no day trend.

In conclusion, the above analysis supports the assumption that the average does not vary over study days for both in all blocks.

\section{An illustrative example of why block averages do not sum to the daily average}
\label{sec:daily-avg-less-block-avg}
\begin{figure}
    \centering
    \includegraphics[width=0.7\linewidth]{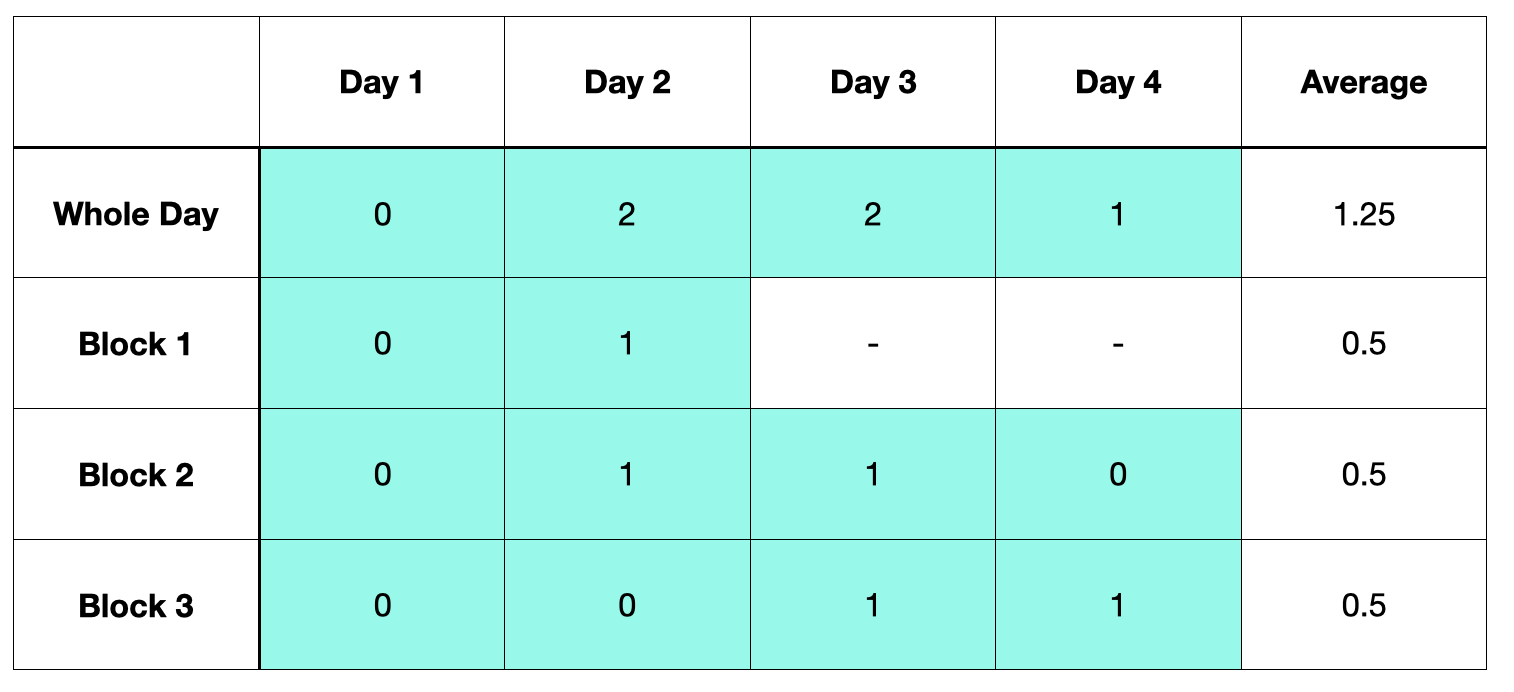}
    \caption{This example illustrates why block goals (average of 0.5 treatments per block) are achieved, but the daily goal (average of 1.5 treatments per day) is not. Blue boxes represent blocks or days with at least one available time, and the numbers inside them show the total number of treatments in each block or day. On day 3 and 4, there is no available time in block 1, so the average for block 1 does not include these two days. However, these two days are included in the daily average calculation because there are available times in blocks 2 and 3.}
    \label{fig:daily-avg-less-block-avg}
\end{figure}

The phenomenon where the block goals of 0.5 treatment per block are achieved, but the daily goal of 1.5 treatments per day is not met, occurs because a participant may be available in one block but unavailable in the other two blocks on a given day. Consequently, the number of days counted for the daily average is greater than the number of days counted for the block averages, resulting in a lower daily average compared to the block averages.
Figure \ref{fig:daily-avg-less-block-avg} provides an example illustrating this scenario. For instance, a participant with four study days may be available in all four evening blocks but only available in two morning blocks. When calculating the average treatment in block 1, the denominator (the number of available block 1s) is 2. However, when calculating the average treatment per day, the denominator (the number of available days) is 4.

\section{Discussion on Not Using KL Divergence for Uniformity}
\label{sec:discuss-no-KL}

In past work, \cite{liao2018just} proposed evaluating uniformity using Kullback-Leibler (KL) divergence between SeqRTS probabilities and the oracle probabilities (Equation \ref{eq:oracle}) at each risk time, then averaging the distance over all risk times in the block. For a day, they use  
\begin{equation}
V_d^{KL} = \dfrac{\sum_{t: X_{d,t}=1} KL(p_{d,t},p_{d,t}^{Oracle}) }{ N_{d}}
\end{equation}

where $KL(p,q) = p \log\left(\frac{p}{q}\right) + (1-p) \log\left(\frac{1-p}{1-q}\right)$, and $ p_t^{Oracle} = 0.5 / N_{d} $.

There are two issues with using KL divergence as a uniformity metric. First, KL divergence is asymmetric, meaning that \(KL(p, q)\) and \(KL(q, p)\) are generally different. One may use Hellinger distance to address the asymmetry issue. The Hellinger distance, given by \(H(p,q) = \frac{1}{\sqrt{2}} \sqrt{(\sqrt{p} - \sqrt{q})^2 + (\sqrt{1-p} - \sqrt{1-q})^2}\), measures how one probability distribution \(Q\) is different from a second, reference probability distribution \(P\). The corresponding uniformity metric \(V_d^{Hell}\) is defined by
\begin{equation}
V_d^{Hell} = \dfrac{\sum_{t: X_{d,t}=1} H(p_{d,t}, p_{d,t}^{Oracle}) }{ N_{d}}.
\end{equation}

Second, uniformity is achieved when \(V_d^{KL}\) and \(V_d^{Hell}\) are zero, but \(V_d^{KL} \neq 0\) (respectively, \(V_d^{Hell} \neq 0\)) even when all treatment probabilities are equal because both KL divergence and Hellinger distance encode both the uniformity and average treatment constraints. We have shown this using an example in Section 4.2 of the main paper. Here we provide a mathematical derivation.

The goal is to show $V_d^{Hell}$ contains the difference in adhering average 0.5 per block, which we denote as $d(p^{\text{SeqRTS}}, p^{\text{Unif}})$. Note that 
$$d(p^{\text{SeqRTS}}, p^{\text{Unif}}) = |(\sum_t p_{d,t}^{\text{SeqRTS}} - 0.5) - (\sum_t p_{d,t}^{\text{Unif}} - 0.5)| = |\sum_t (p_{d,t}^{\text{SeqRTS}} - p_{d,t}^{\text{Unif}})|$$ since  $\sum_t p_{d,t}^{\text{Unif}} =0.5 $

We show $V_d^{Hell} \geq C d(p^{\text{SeqRTS}}, p^{\text{Unif}}) $ where $C > 0$ is a positive constant.

Fix a $t$ where $I_t = B_t = 1$. Denote $p_{d,t}^{\text{SeqRTS}} = p$ and $p_{d,t}^{\text{Unif}} = q$, we have 
\begin{align*}
    H(p,q) =& \frac{1}{\sqrt{2}} \sqrt{(\sqrt{p} - \sqrt{q})^2 + (\sqrt{1-p} - \sqrt{1-q})^2}\\
    \geq &  \frac{1}{2 \sqrt{2}}  (|p-q| + |(1-p) - (1-q)| ) \text{ by Holder's Inequality} \\
    =& \frac{1}{\sqrt{2}} |p-q| \\
\end{align*}
i.e. $\sqrt{2}  H(p,q)  \geq |p-q|$

Plug in $p_{d,t}^{\text{SeqRTS}} = p$ and $p_{d,t}^{\text{Unif}} = q$, and sum over all time, we have
\begin{align*}
    V_{d}^{Hell} =& \frac{1}{N_{d}} \sum_{t = 1,2,..,T,  \mathbf{1}(I_t=1, B_t=1)}H(p_{d,t}^{\text{Unif}},p_{d,t}^{\text{SeqRTS}})\\
    \geq&  \frac{1}{N_{d}} \sum_{t = 1,2,..,T,  \mathbf{1}(I_t=1, B_t=1)}  \frac{1}{\sqrt{2}} |p_{d,t}^{\text{Unif}} - p_{d,t}^{\text{SeqRTS}}| \\
    \geq&  \frac{1}{N_{d}} \frac{1}{\sqrt{2}}  |\sum_{t = 1,2,..,T,  \mathbf{1}(I_t=1, B_t=1)}  (p_{d,t}^{\text{SeqRTS}} - p_{d,t}^{\text{Unif}})| \text{ by Triangle Inequality} \\
    =&  \frac{1}{N_{d}} \frac{1}{\sqrt{2}}   d(p^{\text{SeqRTS}}, p^{\text{Unif}}) 
\end{align*}

Since $N_{d} \leq 48$, we can take $C =  \frac{1}{48\sqrt{2}}$

\section{Details on managing baseline variables}
\subsection{Details on categorizing baseline sedentariness and baseline activeness}
\label{sec:bsl-categorize}

The categorization of step counts within the specific 12-hour window of 8 am to 8 pm offers a nuanced perspective into an individual's activity patterns during standard waking hours. Here we discuss how we categorize activity level using the median daily total number of steps during the baseline period and categorize sedentary behavior using the hours a participant typically spends sitting on a weekday in the past 7 days.

The categorizations of activity level are as follows:
\begin{itemize}
    \item $\leq$ 5,000 steps: Suggestion of low activity level. This aligns with findings that associate fewer than 5,000 steps a day with a sedentary lifestyle \citep{tudor2004many}.
    \item 5,001 to 7,500 steps: A representation of moderate activity level, also suggested by \cite{tudor2004many}. 
    \item 7,501 to 9,000 steps: Suggestion of high activity level. The upper limit of 9,000, rather than the often-quoted 10,000 in behavioural literature, is chosen with consideration of the dataset, where the 3rd quartile is observed at 8,745 steps. This suggests that a significant proportion of the population becomes more active, yet may not consistently hit the well-publicized 10,000 step mark within these 12 hours.
    \item More than 9,000 steps: Reflects individuals who exhibit very high activity levels during this period, often going beyond typical norms.
\end{itemize}

The categorization of sedentary behaviour is as follows:
\begin{itemize}
    \item $\leq$ 4 hours: This range represents individuals who manifest low sedentary behavior during weekdays. Limiting sitting to 4 or fewer hours showcases a lifestyle that incorporates a significant amount of movement or standing activities. Such behavior is align with recommendations from health experts that emphasize the importance of regular movement throughout the day \citep{biswas2015sedentary}.
    \item 4.01 to 8 hours: This category suggests a moderate level of sedentary behavior. It is a notable range because it encompasses the typical workday duration for many office jobs. Individuals in this bracket likely have occupations that necessitate prolonged sitting or are in environments where movement is limited during the bulk of the day.
    \item More than 8 hours: Individuals in this range exhibit a high sedentary lifestyle. Daily sitting durations exceeding 8 hours can exponentially amplify the risks associated with sedentary lifestyles, including chronic diseases and other health issues \citep{owen2010too}.
\end{itemize}

\subsection{Detail of Addressing Missingness in Baseline Covariates}
\label{sec:bsl-missingness}
Addressing missingness in the candidate baseline covariates, specifically the \texttt{sedentary behavior} values absent for 9 users, is crucial for analysis stability. We employed three distinct methods, each resulting in a separate dataset for further evaluation. The missing \texttt{sedentary behavior} data was managed through: (i) K-Nearest Neighbor (KNN) imputation which gave rise to the \texttt{df\_KNN} dataset, (ii) median imputation which leads to the \texttt{df\_median} dataset, and (iii) a straightforward approach where entries with missing data were omitted, producing the \texttt{df\_drop} dataset. We did not employ multiple imputation due to its potential drawbacks in this context: One primary concern is its complexity, which might not necessarily yield better results compared to simpler methods given the size of our dataset. Multiple imputation might introduce additional uncertainty and does not necessarily ensure enhanced analysis stability. Evaluating our methodologies on these three datasets sufficiently allows us to ensure the robustness and stability of our analysis.

\section{Ensure No Endogeneity from the Time-Varying Covariates}

\subsection{Imputation of Partial Risk Indicator}
\label{appendix:imputation-I_t}
Endogeneity arises when an independent variable in a model is correlated with the error term. It is known that endogeneity can affect GEE estimates because it introduces correlation between the predictors and the error term, leading to biased and inconsistent estimates (\citeay{diggle2002analysis}). In our paper, caution should be put on the risk time indicator. Specifically, as explained in Section \ref{sec:HSV2V3-trial}, a decision time becomes unavailable if the participant has received any type of notification (including an anti-sedentary message) within the preceding hour. This means the risk time is influenced by the outcome. 

The perfect indication of a participant's sedentary behavior, while wearing the watch, can be achieved by eliminating any unavailability of notifications within one hour and adhering to the other three availability criteria. The partial availability indicator, which we denote as $\tilde I_t$, is one when only the first three availability criteria are satisfied. Recall that each availability criterion is defined by the converse of its corresponding unavailability criterion. The partial risk indicator $\tilde X_t := \mathbf{1}(B_t = 1, \tilde I_t = 1)$ provides a comprehensive insight into a participant's behavior, without any interference from the algorithm's decision.

Regrettably, $\tilde I_t$ were not maintained during the study, except for a brief period at the beginning. Consequently, we are mostly unable to obtain this ideal indication of a participant's sedentary behavior. To handle this issue, we devised two different strategies to approximate it and opted for the one that minimized the prediction error during the brief period when the indicators for the other three availability criteria were available. The two methods are: 
\begin{enumerate}
	\item \textit{Estimation Method 1}: Impute $\tilde I_t = 1$ for all decision times within one hour following the treatment. This is based on the assumption that there is no treatment effect and participants are likely to be sedentary and available within the treated hour, since they have already been sedentary and available.
	
	\item  \textit{Estimation Method 2}: Impute \(\tilde I_t = B_t\). This is based on the assumption that \(B_t = 1\) implies availability on the other three criteria, and \(B_t = 0\) means the user is not available (i.e., did not wear the watch, was being active, or the watch lost connection).

\end{enumerate}

The indicators of the other three unavailabilities were available before 2019-11-10 (1,555 participant days). These were used as our test set. We calculate the prediction error, defined as the proportion of correct predictions on the test set. Estimation method 1 gives an error of 0.10, and estimation method 2 gives an error of 0.15. Hence, we proceed with estimation method 1.

\subsection{Effect of the Treatment on Sedentariness}
\label{sec:CEE-treatment-sedentariness}
In this section, we provide data analysis to support that messages sent on one day do not impact sedentariness on the next day. We assess the causal effect of whether there was one treatment sent on the previous day \(d-1\) on today's sedentariness \(N_d\) (number of risk times on day \(d\)).

Denote by $Z_d$ the vector of daily characteristic collected after day $d-1$ and just before the randomization happened on day $d$. $Z_0$ represent the baseline covariates. Recall the $Y_d$ represents the number of anti-sedentary messages sent on day $d$, $I_d$ represents the availability of day $d$. The treatment variable we are interested is $A_d = \mathbf{1}(Y_d > 0)$, whether or not there is one treatment sent in the previous day $d-1$. The outcome of interest is $N_d$, number of risk times (aka the sedentariness) on day \(d\).   Data arrives in a sequence of 
$$\{Z_0, A_0 = 0;  Z_1, I_1, Y_1,A_1, Z_2, I_2, Y_2, A_2; ..., Z_{d-1}, I_{d-1}, Y_{d-1}, A_{d-1}; Z_d, I_d, Y_d, A_d, ..., Z_D, I_D, Y_D\}$$ 

We denote the information occured up to day $d$ as  $H_d= (\bar Z_d, \bar A_{d-1}, \bar I_d)$. In this analysis we assume participants are independent and identically distributed from an unknown distribution $P_0$. 

To define treatment effects, we use the potential outcomes framework \citep{robins1986new, rubin1974estimating}. For a participant, the potential outcomes are 
$$\{ N_1, A_1, N_2(a_1), A_2(a_1), ..., N_{D-1} (\bar a_{D-2} )  ,  A_{D-1}(\bar a_{D-2} ), N_D(\bar a_{D-1} ) \}$$

We assess the the marginal causal excursion effect (MCEE, \citep{boruvka2018assessing}) of $A_d$ on $N_d$:   
\begin{equation}
\begin{split}
    \beta = &E\{ E[N_d(\bar A_{d-2}, 1) | H_d, I_{d-1} = 1] | I_{d-1} = 1\} - \\
    & E\{ E[N_d(\bar A_{d-2}, 0) | H_d, I_{d-1} = 1] |  I_{d-1} = 1\} 
\end{split}
\end{equation}\footnote{Here by $H_d, I_{d-1} = 1$ we mean $H_d / I_d, I_{d-1} = 1$. We use the former notation for conciseness.}
The MCEE represents the contrast between at least one push notification ($a_{d-1} = 1$) versus no push notification ($a_{d-1} = 0$) on day $d-1$, following the observed treatment history up to day $d-2$. The quantity marginalizes on all kind of participants and day. 

To represent the CEE in terms of the observed quantities, we make the following identification assumptions 

\textit{Assumption 1 (Consistency)} $N_d(\bar A_{d-1}) = N_d$,  for all $d= 1,..., D$. $A_d(\bar A_{d-1}) = A_d$,  for all $d= 1,..., D-1$.

\textit{Assumption 2 (Positivity)} If $P(H_d = h_d, I_{d-1} = 1) > 0$, then $P(A_d = a| H_d = h_d, I_{d-1} = 1) >0$ for $a \in \{0, 1\}$. 

\textit{Assumption 3 (Sequential ignorability)} For $d = 1,..., D$, $N_d(\bar a_{d}) \indep A_{d} | H_d$ 

With the assumptions, the CEE is identified as 
\begin{equation}
\begin{split}
    \beta =  &E\{ E[N_d | H_d, I_{d-1} = 1, A_{d-1} = 1] |  I_{d-1} = 1\} - \\
    &E\{ E[N_d | H_d, I_{d-1} = 1, A_{d-1} = 0] | I_{d-1} = 1\} 
\end{split}
\end{equation}

To estimate the causal effect, we need a working model for \(E\{ E[N_d | H_d, I_{d-1} = 1, A_{d-1} = 0] | I_{d-1} = 1\}\). A working model approximates the true conditional expectation as well as possible using variables that predict the outcome but not the treatment (called control variables), though it does not need to be perfectly specified. A good working model will improve the estimation efficiency of the effect. We choose \(\alpha \cdot Z_d\) for the working model, where \(Z_d\) is the past 5-day average of the number of sedentary times prior to day \(d\), as we believe \(Z_d\) predicts \(N_d\), and the model shows that \(Z_d\) is significant.

The effect estimate is 0.231 with a p-value of 0.074, indicating no treatment effect. We conclude that there is no effect of treatment on the next day's sedentary behavior, i.e., participants do not respond to the intervention by becoming less sedentary the next day.

\section{The Robustness of the Algorithm's Performance}
\label{sec:robustness-performance}

One factor complicating real-world decision-making is participants' unavailability during a one-hour period after each treatment, as mentioned in Section \ref{sec:HSV2V3-trial}. Due to this criterion, even oracle sampling probabilities (see Equation \ref{eq:oracle}) may not fulfill the average treatment constraint. This is because the oracle only knows the risk times but not the actual treatment assignments, which are determined by a Bernoulli draw. Consider a block where a user is sedentary for 5 times. The oracle probability is set to \(0.5 / 5 = 0.1\). If a treatment is sent on the third instance, then the fourth and fifth instances would not be allowed to randomize, i.e., the probability will be forced to be 0 instead of 0.1 if these instances fall within the hour after the treatment. In general, a treatment decreases the opportunities for randomization. As a result, the actual average treatment delivered is less than desired. Thus, the oracle approach, which distributes a budget of 0.5 evenly across all risk times in probability, could result in the actual average treatment being lower than 0.5. This is verified in a simulation below.

Though it affects the oracle approach, this real-world complication was addressed by the study designers during the parameter tuning phase of SeqRTS. One way this was handled is by setting \(\hat{N}_0\), the tuning parameter of the average treatment goal, to be much larger than the actual goal of 0.5 (recall in Section 3 that \(\hat{N}_0\) was set to $1.8$).

The above hypothesizes that the oracle fails to achieve the target of 0.5 treatments per block, but SeqRTS is capable of meeting this goal. We now use a simulation to demonstrate this. In the simulation, we go through all available participant blocks in HSV2V3 (refer to Section 4 for the selection process). It is important to note that the availability indicator is dependent on the SeqRTS algorithm due to the one-hour unavailability following treatment. We eliminated this dependence by using an availability indicator based on other criteria, as detailed in Appendix \ref{appendix:imputation-I_t}. To ensure a fair comparison, we also simulated the performance of SeqRTS instead of using the actual randomization probabilities and treatments from the study.

\begin{figure}
	\centering
	\includegraphics[width=0.95\linewidth]{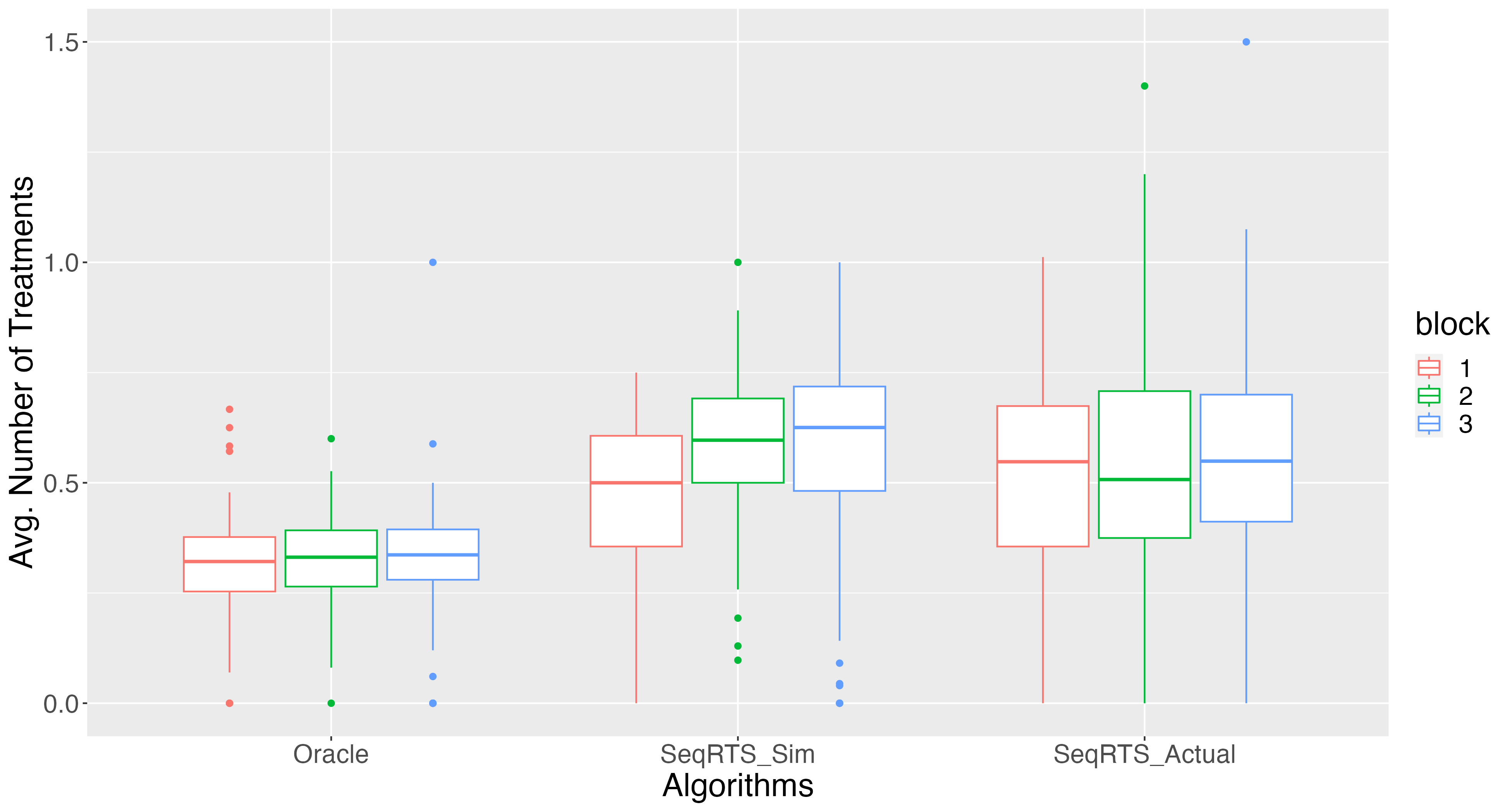}
	\caption{Boxplot of the average number of treatments per participant for each block calculated using SeqRTS and the Oracle. SeqRTS\_Sim represents the performance of SeqRTS on the simulated risk times, while SeqRTS\_Actual represents the performance of SeqRTS on actual decision times in HSV2V3.}
	\label{fig:SeqRTS-boot-vs-actual-uniform-avg-trt}
\end{figure}

The results of the simulation comparing SeqRTS and uniform sampling are presented in Figure \ref{fig:SeqRTS-boot-vs-actual-uniform-avg-trt}. Indeed, the oracle yielded an average of fewer than 0.5 treatments per block, while the SeqRTS algorithm achieved close to an average of 0.5 treatments per block. We also presented the average treatment calculated using the actual data in HSV2V3. The performance of the simulated SeqRTS differs from that of the actual one because, in the simulation, there are more risk times (due to the imputation of the risk times in Appendix \ref{appendix:imputation-I_t}) than actually occurred. Overall, this comparison demonstrates that SeqRTS is robust in real-world environments.

\end{document}